\documentclass[usenatbib]{mnras}

\usepackage{hyperref}

\usepackage{newtxtext,newtxmath}

\usepackage[T1]{fontenc}
\usepackage{ae,aecompl}

  \usepackage{amsmath}
\usepackage{enumitem}
\usepackage{enumerate}
\usepackage{float}
 \usepackage{multirow}
  \usepackage{longtable}
  \usepackage{lipsum}
  \usepackage{gensymb}
  \usepackage{chngcntr}
   \usepackage{multirow}

\usepackage{graphicx}

\usepackage{float}
  \usepackage{multirow}
\usepackage[toc,title,page]{appendix}

\title[The QN  in FRB data?]{The Quark-Nova model for FRBs: model comparison with observational data}

\author[Ouyed et al.]{
Rachid Ouyed,$^{1}$\thanks{E-mail: rouyed@ucalgary.ca}
Denis Leahy,$^{1}$
and Nico Koning$^{1}$
\\
$^{1}$Department of Physics and Astronomy, University of Calgary, 2500 University Drive NW, Calgary, AB, T2N 1N4, Canada
}

\date{Accepted: 2025 January 27. Received: 2025 January 27; in original form 2024 August 7}

\pubyear{2021}

\hypersetup{draft}    
\begin{document}
\label{firstpage}
\pagerange{\pageref{firstpage}--\pageref{lastpage}}
\maketitle

\begin{abstract}
We utilize the Quark-Novae (QN) model for Fast Radio Bursts (FRBs; \citealt{ouyed2021}) to evaluate its performance in reproducing the distribution and statistical properties of key observations. These include frequency, duration, fluence, dispersion measure (DM), and other relevant features such as repetition, periodic activity window, and the sad trombone effect. In our model, FRBs are attributed to coherent synchrotron emission (CSE) originating from collisionless QN chunks that traverse ionized media both within and outside their host galaxies. By considering burst repetition from a single chunk and accounting for the intrinsic DM of the chunks, we find agreement between our model and the observed properties of FRBs. This agreement enhances our confidence in the model's effectiveness for interpreting FRB observations.
Our model generates testable predictions, allowing for future experiments and observations to validate and further refine our understanding of FRBs.
\end{abstract}


\section{Introduction}
\label{sec:intro}

Fast Radio Bursts (FRBs) are intense and transient bursts of cosmological radio waves first discovered in 2007 (\citealt{lorimer2007}; see also \citealt{thornton2013}). The FRB pulse duration spans orders of magnitude  from micro-seconds 
 to seconds (e.g. \citealt{snelders2023}), a frequency ranging from 110 MHz to 8 GHz (\citealt{gajjar2018,pleunis2020})
  and appear to be a band-limited phenomenon (e.g. \citealt{pearlman2020,kumar2021}). 
 Their redshifts range from $\sim 0$ to $z>1$ and they 
  exhibit dispersion measures (DM) of hundreds of pc cm$^{-3}$ (\citealt{cordes2019})
  with typical uncertainty of $\sim 1$ pc cm$^{-3}$.  At cosmological distances their estimated radio energies are  
  $10^{38}$-$10^{40}$ ergs (e.g. \citealt{petroff2019}). 
  Some FRBs are highly linearly polarized (e.g. \citealt{petroff2019}) while a small fraction show
significant circular polarization (e.g. \citealt{masui2015}).

Over 600 FRB events have been identified with a small percentage of them classified as repeaters  (e.g. \citealt{chime2021,spitler2016,fronseca2020,chime2023}).  
 FRB 180916.J0158$+$65 shows  a $\sim 16.35$ day periodic activity window ($\sim$ 5 day wide)  (\citealt{marcote2020}). 
The repeating and non-repeating FRBs seem to show statistically different dynamic spectra.
Repeaters are associated with wider bursts, narrower band spectra and show the  ``sad trombone" (multiple sub-bursts) effect  (\citealt{pleunis2021,chime2021}). Also reported, is a statistical difference in  the DM and extragalactic DM 
distributions between repeating and apparently non-repeating sources, with repeaters having a lower mean DM   (\citealt{andersen2023}).  

Precise localization has  associated a dozen FRBs to host galaxies suggesting a diverse range of sources and regions from which these bursts originate while other FRBs are spatially offset from their hosts  (\citealt{bannister2019,prochaska2019a,mannings2021,bhandari2022,mckinven2023}).
The variety of galaxy associations  (\citealt{prochaska2019b}), 
 selection biases (\citealt{macquart2018}) and propagation effects  (e.g. \citealt{petroff2022}) makes it hard to pinpoint 
 the  properties of an FRB progenitor.  This has allowed for a plethora of models for FRBs
 which  include 
 the merger of compact objects (\citealt{metzger2019}), magnetars  (\citealt{kaspi2017}; see also \citealt{cordes2019})
 flares from active stars (\citealt{loeb2014}) and pulsars (\citealt{bera2019,connor2016}). 
  The windowed activity seen in FRB 180916.J0158$+$65  has been attributed to different mechanisms including binary effects (e.g. \citealt{lyutikov2020}) and precession (e.g. \citealt{li2021}).  A  review of models and different emission mechanisms can be found in 
   \citet{platts2019} and \cite{zhang2023} and references therein.
    Models involving magnetars have support following the 
 confirmation of a Galactic magnetar (SGR 1935$+$2154) producing FRBs (\citealt{rane2020})
 albeit a very dim one.

  FRBs remain an enigma  (e.g. \citet{cordes2019,petroff2022} for recent reviews) which leaves room for new ideas to be explored.
     Here, we explore further our model for FRBs based on the Quark-Nova (QN) model as presented in detail in \citet{ouyed2021} (hereafter, first paper).
A QN is a hypothetical transition of 
an Neutron star (NS) to a quark star (QS) following quark deconfinement in the core of a NS induced by
spin-down or mass-accretion onto the NS (\citealt{ouyed2022micro,ouyed2022macro} and references therein). 
The high brightness temperatures of FRBs point to coherent emission from a compact source with high energy density,
and for this reason many models have invoked  compact stars. Here, we argue instead that FRBs can emerge from
relativistic, low-density objects (which are fragments of the QN ejecta) emitting coherent synchrotron emission (CSE);
the FRB in our model. 

\begin{figure*} 
\centering
\includegraphics[scale=0.3]{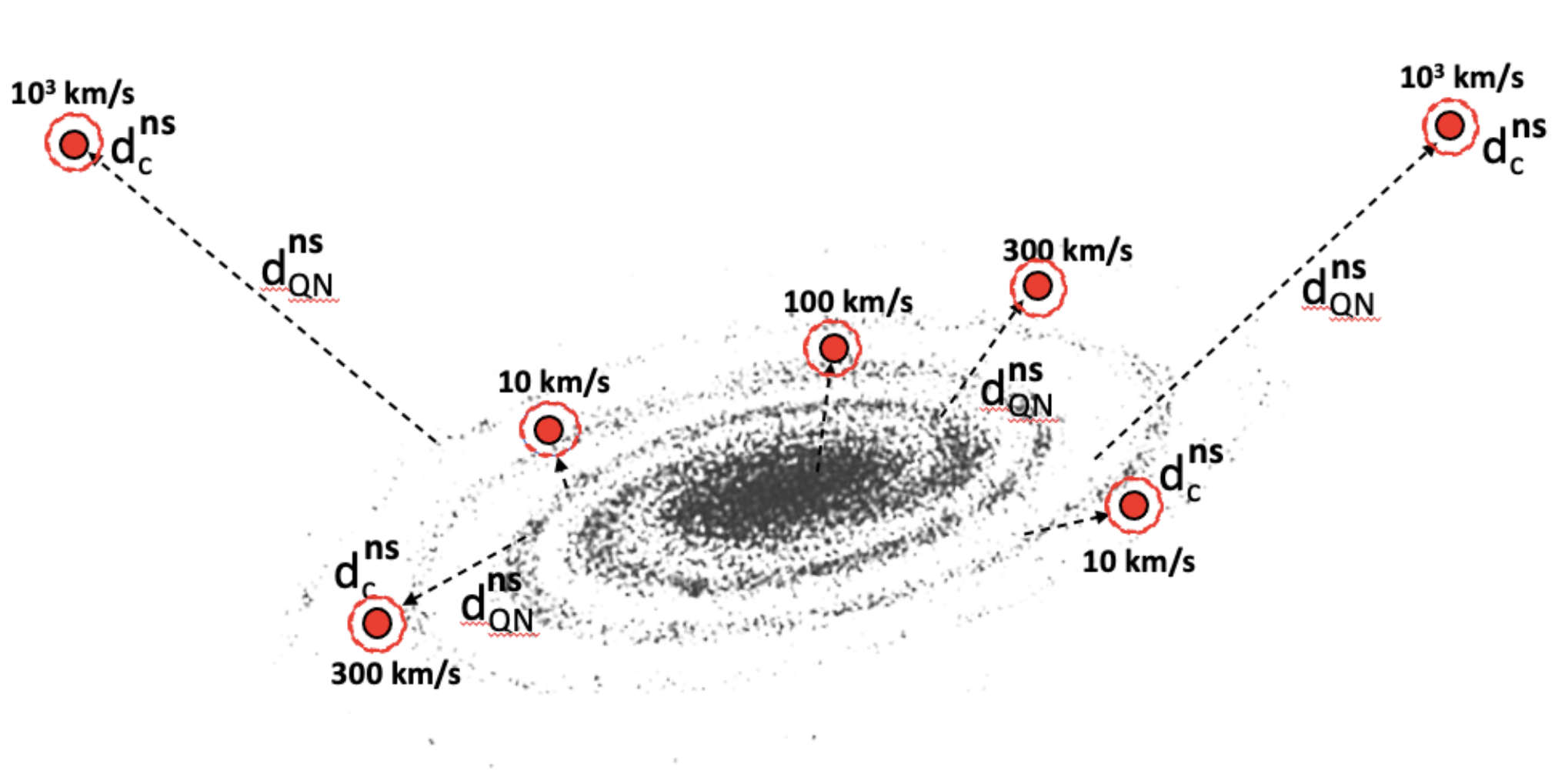}
\includegraphics[scale=0.5]{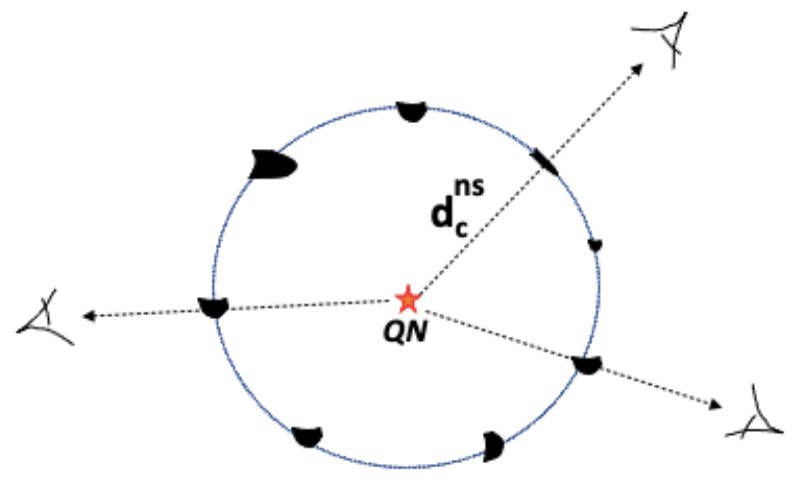}
\caption{{\bf Left panel}: There are two time delays in our model: (i) the time between the birth of the NS (the supernova; SN)
 and when the NS experiences the QN event ($\tau_{\rm QN}\sim 10^7$ years) during
 which time the NS would have travelled a distance $d_{\rm QN}^{\rm ns}$ ($\sim$ kpcs; see Eq. (\ref{eq:dQNns}));
 (ii) the time it takes the QN chunks to become collisionless ($\tau_{\rm c}^{\rm ns}\sim 10^2$ years; see Eq. (\ref{eq:tcns})) during which time a chunk would have travelled
 a distance $d_{\rm c}^{\rm ns}<<d_{\rm QN}^{\rm ns}$ from the QN site (see Eq. (\ref{eq:dcns})).  The velocities shown in the left panel are the NSs kick velocities $v_{\rm kick}$. The red dot portrays the NS location in the galaxy when it experiences the QN event while the red circle around it shows the distance $d_{\rm c}^{\rm ns}$   travelled by the  QN chunks until they become collisonless.  {\bf Right panel}:  The collisionless QN chunks  at $d_{\rm c}^{\rm ns}$ emitting FRBs as CSE (see Appendix \ref{appendix:CSE}) as seen by different observers. 
}
\label{fig:single-chunk}
\end{figure*}

\begin{figure*} 
\centering
\includegraphics[scale=0.36]{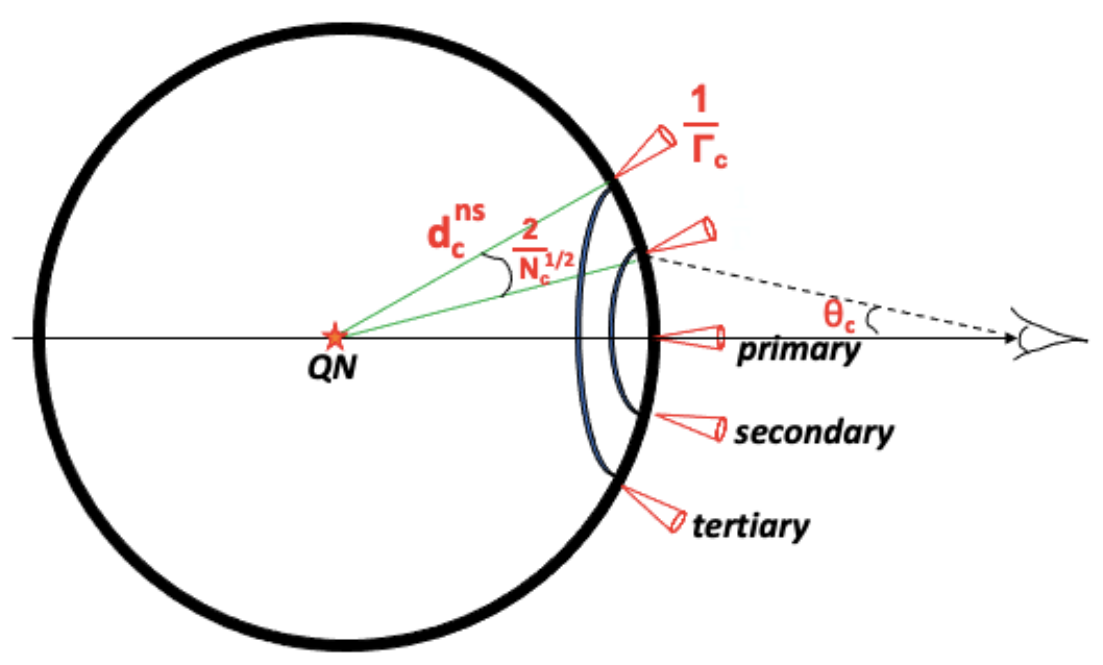}
\hskip 0.2in
\includegraphics[scale=0.4]{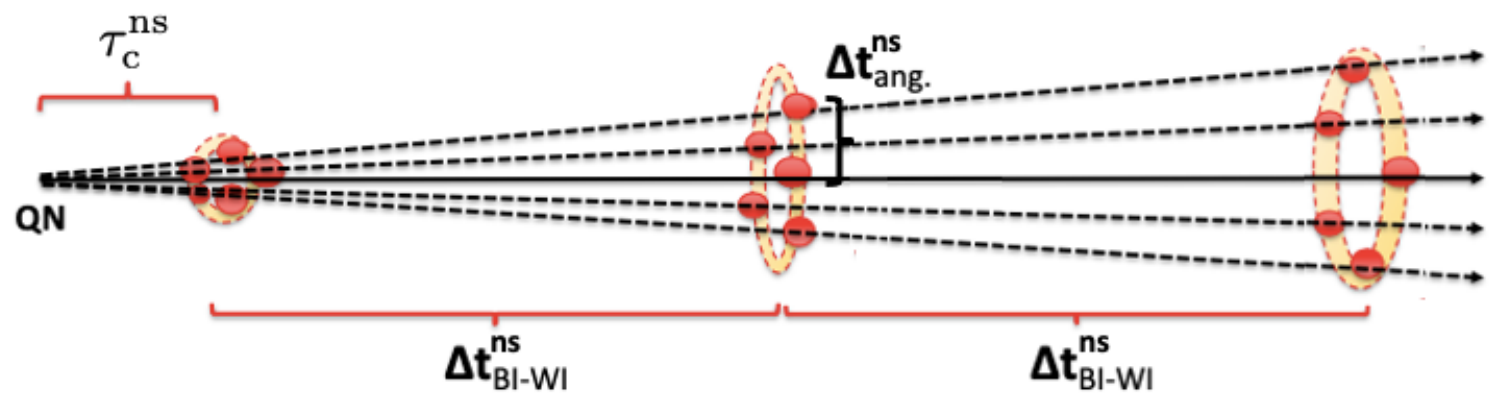}
\caption{  {\bf Left panel}: Multiple (the primary, the secondary and the tertiary) collisionless chunks at $d_{\rm c}^{\rm ns}$ from a QN at a viewing angle $\theta_{\rm c}$. 
The typical angular separation between adjacent chunks is $2/N_{\rm c}^{1/2}$ where $N_{\rm c}$ is the total number of chunks ejected by the QN (see Appendix \ref{appendix:angular-periodicity}). 
The average angular time separation between successive chunks is $ \Delta t_{\rm ang.}^{\rm obs.}\sim $ hours (see Eq.(\ref{eq:Pcang})).
{\bf Right panel}: The first time chunks emits CSE is at 
 $\tau_{\rm c}^{\rm ns}$ which is decades after the QN event. Radial repetition in our model is due to a QN chunk experiencing multiple BI-WI episodes as it travels radially
away from the QN site (see Appendix \ref{appendix:radial-periodicity}). This defines a third distance in our model $\Delta r_{\rm BI-WI}^{\rm ns}$  which is the distance travelled by the chunk between CSE episodes (see Eq. (\ref{eq:distance})) with $d_{\rm QN}^{\rm ns}>>\Delta r_{\rm BI-WI}^{\rm ns}>> d_{\rm c}^{\rm ns}$.
 The time delay between BI-WI episodes, $\Delta t_{\rm BI-WI}^{\rm obs.}$ is given by Eq. (\ref{eq:Pcrad}).}
\label{fig:multiple-chunks}
\end{figure*}

At the heart of our model  is the QN ejecta 
which fragments into millions of chunks as shown in \citet{ouyed2009}.
As the chunks expand away from the QN site (see Figure \ref{fig:single-chunk} and Appendix \ref{appendix:angular-periodicity}) they eventually become collisionless\footnote{When the electron Coulomb collision length inside the chunk is of the order of 
its radius (see Appendix SB in \citealt{ouyed2021}).}  and prone to 
plasma instabilities (once they enter an ionized medium) which yield CSE with properties reminiscent
of FRBs (\citealt{ouyed2021}). Despite being a one-off explosive event, the QN can account for repetition in FRBs
 due to different chunks from the same QN seen at different viewing angles
and at different arrival times (referred to as ``angular repetition" in our model; see Figure \ref{fig:multiple-chunks}).
The Tables in \citealt{ouyed2021} show typical angular repeaters in our model. 
Repeating FRBs have been employed to dismiss the possibility of catastrophic events. However, it does not necessarily mean that a single occurrence cannot produce time-varying FRBs as shown in \citet{ouyed2021} and as we confirm in this paper.

 A NS can increase its core density via spin-down, if born rapidly rotating following a supernova (SN), or
 if it has experienced mass accretion (e.g. from fall-back during the SN). This increases the density in its core 
 (with the onset of quark deconfinement) which triggers the 
  QN before the SN ejecta has dissipated. In particular,  for short delays (i.e. only years) between the SN and the QN,
  the SN ejecta is dense enough that the QN ejecta's kinetic energy is harnessed before the chunks become collisionless 
yielding super-luminous SNe and Gamma Ray Bursts, respectively (\citealt{ouyed2020}). 

 However, most NSs are born slowly rotating with a birth period distribution 
 mean value of $\sim 300$ ms (\citealt{faucher2006}). For these,  
the time it takes for their core to reach quark deconfinement density is in the tens of millions of years; hereafter we adopt a fiducial
value of $\tau_{\rm QN}\sim 10^7$ years (e.g. \citealt{staff2006}). 
I.e. the QN event is separated in time and space from the SN.
The time delay between the QN and when the
chunks become collisionless (and emit FRBs) as they travel in the ambient medium is decades. This  effectively separates the
QN event from the FRB in time and space (see Figure \ref{fig:single-chunk}).

In this second paper, we extend our investigation of the QN FRB model by taking into account:

(i)  What hereafter we refer to as ``radial repetition" where a given chunk bursts multiple times as it travels radially away from the QN (see Figure \ref{fig:multiple-chunks}).  Thus for a given QN, an observer could see simultaneously ``angular repetition" (chunks at different viewing angles CSE bursting) and ``radial repetition" (each chunk yielding multiple bursting episodes);

(ii) The chunk's intrinsic DM and explore its role in differentiating
angular and radial repeaters and its contribution to the extra-galactic DM.

   The paper is organized as follows: In \S \ref{sec:model}, we discuss salient features of our model and  give the relevant equations.
In \S \ref{sec:single-chunk}, we consider and simulate CSE emission from the QN primary chunk (i.e. along the observer line-of-sight)
 and compare it to FRB data. The case of CSE from multiple chunks from the same QN, yielding ``angular+radial" repetition, is presented in \S \ref{sec:multiple-chunks}. We discuss
our findings in \S \ref{sec:discussion} and list some testable prediction before we conclude in \S \ref{sec:conclusion}.

\section{our model}
\label{sec:model}

 Hereafter, the dimensionless quantities are defined as $f_x=f/10^x$ with quantities in cgs units. 
  Unprimed quantities are in the chunk's frame while the superscripts ``ns" and ``obs" refer to the NS frame
  (i.e. the ambient medium) and the observer's frame, respectively.

The energy released from the conversion of a NS of mass
 $M_{\rm NS}$ to a QS  can be estimated as $E_{\rm QN}=(M_{\rm NS}/m_{\rm n})\times E_{\rm conv.}\sim 10^{53}$ ergs where
$E_{\rm conv.}\sim 100$ MeV ($\sim 10^{-4}$ erg) is the energy release per converted neutron of mass $m_{\rm n}$ (e.g. \citealt{weber2005}).  The QN ejecta is the 
NS crust of mass $M_{\rm QN}\sim 10^{-5}M_{\rm NS}$. A small percentage $\zeta_{\rm KE}\sim 10^{-2}$-$10^{-1}$ is converted to the
kinetic energy of the QN ejecta which acquires a 
Lorentz factor $\Gamma_{\rm QN}=  \zeta_{\rm KE} E_{\rm QN}/M_{\rm QN}c^2\sim10^2$-$10^3$. 
The remaining $\sim$ 90\% of the energy released in the conversion to a QS is lost to neutrinos (\citealt{keranen2005}).
The average kinetic energy of the QN ejecta is thus ${(E_{\rm QN}^{\rm ke})}^{\rm ns}\sim 10^{52}$ ergs. The compact remnant, of mass $M_{\rm QS}= (M_{\rm NS}-M_{\rm QN})\simeq M_{\rm NS}$, is a radio-quiet  QS with properties summarized in \citet[and references therein]{ouyed2017a}.

The QN ejecta breaks into $N_{\rm c}$ chunks with a characteristic mass
$m_{\rm c}= M_{\rm QN}/N_{\rm c}\simeq 10^{22.3}\ {\rm gm}/N_{\rm c, 6}$. We take 
$N_{\rm c}=10^6$ as the fiducial value and is a free parameter which varies from QN-to-QN (\citealt{ouyed2009}).  
The chunk's Lorentz factor we set to 
 a fixed value of $\Gamma_{\rm c}=10^{2.5}$ in this work because the ejected NS crust ($M_{\rm QN}\sim 10^{-5} M_{\rm NS}$) and the 
 converted NS mass ($\sim M_{\rm NS}$) vary little between QNe. A typical QN chunk's kinetic energy
is $(E_{\rm c}^{\rm ke})^{\rm ns}=\Gamma_{\rm c}m_{\rm c}c^2\sim 5.7\times 10^{45}\ {\rm ergs}\times \Gamma_{\rm c, 2.5}m_{\rm c, 22.3}$  which far exceeds the average $10^{39}$ erg FRB radio energy. 
 
 The distance travelled for the NS to experience the QN event is 
\begin{equation}
\label{eq:dQNns}
d_{\rm QN}^{\rm ns}=v_{\rm kick}\tau_{\rm QN}\sim 3\ {\rm kpc}\ ,
\end{equation}
 for typical NS kick velocity at birth $v_{\rm kick}\sim 300$ km s$^{-1}$ (e.g. \citealt{faucher2006}) and for $\tau_{\rm QN}\sim 10^7$ years.

Furthermore, a QN chunk moves outward from the parent NS  (i.e. from the QN site) and becomes collisionless after travelling a distance (see Eq. (\ref{eq:dc}) in Appendix \ref{appendix:angular-periodicity})

\begin{equation}
\label{eq:dcns}
d_{\rm c}^{\rm ns}\sim 28.7\ {\rm pc}\times m_{\rm c, 22.3}^{1/5}\Gamma_{\rm c, 2.5}^{-1/5}n_{\rm amb., -1}^{\rm ns\ -3/5}\ ,
\end{equation}
at which point it is prone to plasma instabilities.  
 This distance $d_{\rm c}^{\rm ns}$ is controlled by the density of the ambient environment, $n_{\rm amb.}^{\rm ns}$, where the NS experiences the QN event.
 The higher the $n_{\rm amb.}^{\rm ns}$ the less distance the chunk needs to travel to become collisionless. 
  In the NS frame it takes
  \begin{equation}
  \label{eq:tcns}
  \tau_{\rm c}^{\rm ns}=\frac{d_{\rm c}^{\rm ns}}{c}\sim 93\ {\rm years}\times m_{\rm c, 22.3}^{1/5}\Gamma_{\rm c, 2.5}^{-1/5}n_{\rm amb., -1}^{\rm ns\ -3/5}\ ,
  \end{equation}
  after the QN for chunks to enter the collisionless regime. We note that $d_{\rm c}^{\rm ns}<<d_{\rm QN}^{\rm ns}$ as illustrated in
  Figure \ref{fig:single-chunk}.
 
 The ambient medium can be any ionized environment in the galactic disk
 (e.g. the warm ionized medium; WIM), the ionized galactic halo, the intra-group/intra-cluster medium
 (IGpM) or the inter-galactic medium (IGM).
Hereafter, the fiducial value of the ambient medium density in the NS reference frame we set to $n_{\rm amb.}^{\rm ns}=0.1$ cm$^{-3}$.  It is
 $\Gamma_{\rm c}n_{\rm amb.}^{\rm ns}$ in the chunk's frame. 
 
 The chunk's number density, $n_{\rm c}$ and  radius ($R_{\rm c}=(m_{\rm c}/\frac{4\pi}{3}n_{\rm c}m_{\rm H})^{1/3}$ with
$m_{\rm H}$ the hydrogen mass) when it becomes collisionless were  derived in Appendix SA in \citet{ouyed2021}:

\begin{align}
\label{eq:ncandRc}
n_{\rm c} &\simeq 3.7\times 10^3\ {\rm cm}^{-3}\times m_{\rm c, 22.3}^{1/10} \Gamma_{\rm c, 2.5}^{12/5}  n_{\rm amb, -1}^{\rm ns\ 6/5}\\\nonumber
R_{\rm c}&\simeq 9.3\times 10^{13}\ {\rm cm}\times m_{\rm c, 22.3}^{3/10} \Gamma_{\rm c, 2.5}^{-4/5}  n_{\rm amb, -1}^{\rm ns\ -2/5}\ .
\end{align} 
  
  Once collisionless, the interaction with the ambient plasma protons (the beam)\footnote{In the chunk's fame where the instabilities develop, the ambient plasma is a beam with Lorentz factor $\Gamma_{\rm c}$.} and the chunk's electrons (the background plasma) triggers successively 
the Buneman and thermal Weibel instabilities (hereafter BI and WI) which yield CSE from the QN chunks which are FRBs in our model (see \citealt{ouyed2021}). 
   The BI converts a percentage $\zeta_{\rm BI}$ of the chunk's kinetic energy to electron thermal energy in the direction of motion of the chunk yielding
   a temperature anisotropy which triggers the thermal WI. This anisotropy is defined by the anisotropy parameter $\beta_{\rm WI}= v_{\perp}/v_{\parallel}<1$ 
   with $v_{\perp}$ and $v_{\parallel}$ the electron's speed in the direction perpendicular and parallel to the chunk's direction of motion (see Appendix \ref{appendix:CSE}). 
   The WI acts to remove the anisotropy by first developing current filaments within the chunk and along its direction of motion (see Figure S3 in \citealt{ouyed2021}) before    
   converting the BI-gained thermal energy into magnetic energy and into accelerating the chunk's electrons to relativistic speeds
   (see \citealt{ouyed2021}). 
  
   The Weibel filament's initial radius  (transverse to the direction of motion of the chunk) is $\lambda_{\rm F, min.}=\beta_{\rm WI}^{1/2}\times c/\nu_{\rm p, e}$  with $\nu_{\rm p, e}\simeq 9\times 10^3\ {\rm Hz}\times n_{\rm c, 0}^{1/2}$ the chunk's plasma frequency; $m_{\rm p}$ and $m_{\rm e}$ are the proton's and electron's mass, respectively. The WI filaments merge and grow in size as $\lambda_{\rm F}(t)=\lambda_{\rm F, min.}\times (1+t/t_{\rm m-WI})^{+\delta_{\rm m-WI}}$
    with $t_{\rm m-WI}= \sqrt{m_{\rm p}/m_{\rm e}}\times \zeta_{\rm m-WI}/\nu_{\rm p, e}$ a characteristic timescale
   related to the merging and growth of WI current filaments with $\zeta_{\rm m-WI}\sim 10^5$ the WI filament merging parameter and  $\delta_{\rm m-WI}>0$ the power index.    The subscript ``m-WI" refers to the merging (i.e. non-linear) phase of the WI.
    The filament's radius grows in size until it reaches a maximum size set by the plasma skin depth so that $\lambda_{\rm F, max.}=c/\nu_{\rm p, e}$.

   Electrons bunches of size $\lambda_{\rm b}$  
   form around and along the WI filaments and scale linearly (and grow)
with the WI filaments as $\lambda_{\rm b}(t)=\delta_{\rm b}\lambda_{\rm F}(t)$ with $\delta_{\rm b}<1$; i.e. $\lambda_{\rm b, min.}=\delta_{\rm b}\lambda_{\rm F, min.}$ and $\lambda_{\rm b, max.}=\delta_{\rm b}\lambda_{\rm F, max.}$.
 
 The merging amplifies the WI magnetic field and accelerates the chunks electrons yielding CSE  from the 
  bunches.  The CSE spectrum for a given bunch extends from the peak frequency $ \nu_{\rm CSE, p}(t)=c/\lambda_{\rm b}(t)=c/\delta_{\rm b}\lambda_{\rm F}(t)$ and cuts-off at the chunk's plasma frequency (see Appendix SD in \citealt{ouyed2021}). That is, 
at any given time a bunch emits in the frequency range 
\begin{equation}
\label{eq:nu-drift}
\nu_{\rm p, e}\le \nu_{\rm CSE} \le \nu_{\rm CSE, p}(t)=\nu_{\rm CSE, p, max.}\times \left( 1+\frac{t}{t_{\rm m-WI}}\right)^{-\delta_{\rm m-WI}}\ ,
\end{equation}
where $\nu_{\rm CSE, p, max.}=\nu_{\rm CSE, p}(t=0)$. I.e. 
 $\nu_{\rm CSE, p, max.}= c/\lambda_{\rm b, min.}= c/\delta_{\rm b}\lambda_{\rm F, min.}=\nu_{\rm p, e}/(\delta_{\rm b}\beta_{\rm WI}^{1/2})$.

For a given chunk,  the CSE peak frequency decreases over time from $\nu_{\rm CSE, p, max.}$
   to $\nu_{\rm CSE, p, min.}= c/\lambda_{\rm b, max.}=c/\delta_{\rm b}\lambda_{\rm F, max.}=\nu_{\rm p, e}/\delta_{\rm b}$.  I.e. the emission band narrows down over time from $(\nu_{\rm CSE, p, max.}-\nu_{\rm p, e})/\nu_{\rm p, e}=(1/\delta_{\rm b}\beta_{\rm WI}^{1/2}-1)$ to $(\nu_{\rm CSE, p, min.}-\nu_{\rm p, e})/\nu_{\rm p, e}=(1/\delta_{\rm b}-1)$.
   
   {\it The combined BI-WI tandem effectively converts kinetic energy of the chunk's electrons to CSE (the FRB in our model) until
   the instability saturates and shuts-off.}  As an extension of our first paper, here we consider magnetic field dissipation by Coulomb
   collisions due to trapped chunk electrons at WI saturation which allows for subsequent BI-WI episodes.

  \begin{table*}
  \caption{Model parameter with fiducial values shown.}
\begin{tabular}{|l|l|l|l|}\hline
 \multirow{5}{*}{Astrophysics} &  \multirow{2}{*}{Quark-Nova (QN)} &  $\Gamma_{\rm c}=10^{2.5}$ (QN chunk's Lorentz factor) \\  \cline{4-4}
  & &  $N_{\rm c}=10^6$ (number of chunks per QN) \\\cline{2-4}
  & \multirow{3}{*}{Ambient medium} & $n_{\rm amb.}^{\rm ns}=0.1$ cm$^{-3}$ (QN event environment set by $\tau_{\rm QN}v_{\rm kick}$; neutral or ionized) \\ \cline{4-4}
  & &  $n_{\rm ionized}^{\rm ns}=10^{-3}$ cm$^{-3}$ (BI-WI event environment; ionized)\\ \cline{4-4}
   & &  $L_{\rm ionized}=10$ kpc (size of the ionized medium for the onset of the BI-WI)\\ \cline{4-4}\hline
  \multirow{6}{*}{Plasma Physics} & \multirow{1}{*}{Buneman (BI)} &  $\zeta_{\rm BI}=0.1$ (fraction of chunk's electron energy  converted to thermal anisotropy)\\    \cline{2-4}
  & \multirow{5}{*}{Weibel (WI)} &  $\beta_{\rm WI}=0.1$ (ratio of transverse to longitudinal speed of electrons at the onset of the WI) \\  \cline{4-4}
  & &  $\zeta_{\rm m-WI}=10^5$ (Weibel filament merging timescale parameter)\\
  &  & $\delta_{\rm m-WI}=1.0$ (Weibel filament merging power index)\\
  & &  $\delta_{\rm b}=1.0$ (electron bunch's scaling parameter) \\ 
  & & $\alpha_{\rm diff.}=1.0$ (Weibel magnetic field diffusion parameter) \\ \cline{2-4}\hline
\end{tabular}
\label{Table:AllParameters}
\end{table*}

  \subsection{Model's parameters}
   
   Table \ref{Table:AllParameters} summarizes the parameters in our model. These parameters determine all of the properties of the FRBs except the offset of the NS from its birthplace. The parameters related to the NS offset are $\tau_{\rm QN}$ and $v_{\rm kick}$ which controls the distance
travelled by the NS, $d_{\rm QN}^{\rm ns}$ before it experiences the QN event. I.e. these only enter because they determine $n_{\rm amb.}^{\rm ns}$.
What is relevant to this study are parameters which pertain to the property of the chunk ($\Gamma_{\rm c}, m_{\rm c}, n_{\rm c}$) and 
 the plasma physics parameters related to the BI-WI  ($\zeta_{\rm BI},\beta_{\rm WI},\zeta_{\rm m-WI},\delta_{\rm m-WI},
 \delta_{\rm b}, \alpha_{\rm diff.}$). We have already argued that $\Gamma_{\rm c}$ is not expected to vary much between QNe
 and that $n_{\rm c}\propto n_{\rm amb.}^{\rm ns 6/5}$. This leaves $m_{\rm c}= 10^{-5}M_{\rm NS}/N_{\rm c}$, where $N_{\rm c}$
 is the number of chunks per QN which varies, as the QN-related parameter.

In this study, we take $\delta_{\rm b}=1$ and $\alpha_{\rm diff.}=1$. Setting $\delta_{\rm b}=1$ means that
the electron bunches are of the same size as the WI filament radius. The implication is that 
the CSE peak frequency decreases over time from $\nu_{\rm CSE, p, max.}= \nu_{\rm p, e}/\beta_{\rm WI}^{1/2}$
   to the chunk plasma frequency, $\nu_{\rm CSE, p, min.}=\nu_{\rm p, e}$. 
   The $\alpha_{\rm diff.}=1$ simplification means that we consider the maximum dissipation timescale by electron Coulomb collisions
   once the WI shuts-off and the corresponding magnetic field reached saturation (see Appendix \ref{appendix:radial-periodicity}).
 
 \begin{table*} 
 \centering
\caption{Mean values for parameters explored in this study. We consider a log-normal distribution with $\sigma (\log{x})=1.0$ for all parameters.}
\begin{center}
\begin{tabular}{|c|c|c|c|c|c|c|}\hline
$\bar{z}$& $\bar{m}_{\rm c}\ ({\rm gm})^{*}$  & $\bar{n}_{\rm amb.}^{\rm ns}$ (cm$^{-3})$& $\bar{\zeta}_{\rm BI}$ & $\bar{\beta}_{\rm WI}$ & $\bar{\zeta}_{\rm m-WI}$ &  $\bar{\delta}_{\rm m-WI}$\\\hline 
0.4 & $10^{22.3}$  & $10^{-1}$& $10^{-1}$  & $10^{-1}$ & $10^5$ & 1.0\\\hline
\end{tabular}\\
\end{center}
$^*$ From $\bar{m}_{\rm c}=10^{-5}M_{\rm NS}/N_{\rm c}\sim 10^{22.3}\ {\rm gm}/N_{\rm c, 6}$ with $N_{\rm c}=10^6$ the fiducial value. 
\label{table:fiducial-values}
\end{table*}
   
    Table \ref{table:fiducial-values} lists the parameters we varied in this study and their fiducial mean values.
These consist of the astrophysical parameters 
divided into the QN parameter $N_{\rm c}$ (which sets $m_{\rm c}$) and the ambient density $n_{\rm amb.}^{\rm ns}$ (which gives  $n_{\rm c}$ from Eq. (\ref{eq:ncandRc}) and thus $\nu_{\rm p, e}$). The plasma parameters  are $(\zeta_{\rm BI},\beta_{\rm WI},\zeta_{\rm m-WI},\delta_{\rm m-WI})$,  with $\zeta_{\rm BI}<1$
 the percentage of electron kinetic energy converted to heat by the BI (until BI saturation), $\beta_{\rm WI}<1$ defining the initial size of the
 electron bunch $\lambda_{\rm b, min.}=\lambda_{\rm F, min.}$. 
  The CSE peak frequency is then $\nu_{\rm CSE, p}(t)=c/\lambda_{\rm b}$(t) which decreases over time
  as the bunches grow in size. The parameter $\zeta_{\rm m-WI}$  sets
  the filament (and thus bunch) merging timescale which effectively defines the  CSE characteristic duration
  while $\delta_{\rm m-WI}$ is the power index for peak frequency drift as given in Eq. (\ref{eq:nu-drift}) (see also Eq. (\ref{eq:nu-drift-appendix}).
  
   Variation in astrophysical parameters in our model is embedded in the distribution of chunk mass (for a given QN) and ambient plasma density
  which portrays the different environments in which a QN can occur. Stochasticity in our model is inherent to 
  the distribution of the plasma parameters.  The CSE frequency, duration and other relevant quantities
 can all be expressed in terms of the chunk's plasma frequency as shown 
   in Appendix \ref{appendix:CSE}. 

\subsection{Observer's frame}
\label{sec:observerframe}

   In the observer's frame, the frequency and time  are $\nu^{\rm obs.} = \frac{D(\Gamma_{\rm c},\theta_{\rm c})}{(1+z)}\nu$
and $t^{\rm obs.} = \frac{(1+z)}{D(\Gamma_{\rm c},\theta_{\rm c})} t$.  
The Doppler factor is $D(\Gamma_{\rm c},\theta_{\rm c})\simeq 2\Gamma_{\rm c}/f(\Gamma_{\rm c},\theta_{\rm c})$ with 
$f(\Gamma_{\rm c},\theta_{\rm c})=1+(\Gamma_{\rm c}\theta_{\rm c})^2$ in the $\Gamma_{\rm c}^2>>1$ and $\theta_{\rm c}<<1$ limit; $\theta_{\rm c}$ is the viewing angle
with respect to the observer's line-of-sight (l.o.s.; see Figure \ref{fig:multiple-chunks}). 
The fluence we calculate from $\mathcal{F}=\frac{D(\Gamma_{\rm c},\theta_{\rm c})^3 E_{\rm CSE}}{4\pi d_{\rm L}^2 (1+z)\nu_{\rm CSE, p, max}^{\rm obs.}}$
where $d_{\rm L}$ is the luminosity distance and $z$ is the redshift; we assume a flat spectrum. Here, $E_{\rm CSE}\propto \nu_{\rm p, e}^{-2/3}$ is the
energy released as CSE per BI-WI episode given  by Eq. (\ref{eq:chunk-frame}) in Appendix \ref{appendix:CSE}
with $E_{\rm CSE}<< (E_{\rm c}^{\rm ke})^{\rm ns}=\Gamma_{\rm c}m_{\rm c}c^2$. 

The chunk's intrinsic DM  as measured by the observer is DM$_{\rm c}^{\rm obs.}= \frac{D(\Gamma_{\rm c},\theta_{\rm c})}{(1+z)} n_{\rm c}R_{\rm c}$. It is expressed in terms of the plasma frequency by using   $R_{\rm c}\propto (m_{\rm c}/n_{\rm c})^{1/3}$
from Eq. (\ref{eq:ncandRc}) and $n_{\rm c}\propto \nu_{\rm p, e}^2$.  

       For a given chunk of mass $m_{\rm c}$ and Lorentz factor $\Gamma_{\rm c}$, the 
ambient plasma density $n_{\rm amb.}^{\rm ns}$ gives us the chunk's number density $n_{\rm c}$ and thus
its plasma frequency $\nu_{\rm p, e}$ when it becomes collisionless.  I.e.
\begin{align}
\label{eq:plasma-frequency}
\nu_{\rm p, e}^{\rm obs.}
&\simeq 0.35\ {\rm GHz}\times \frac{1}{(1+z)f(\Gamma_{\rm c},\theta_{\rm c})}\times  m_{\rm c, 22.3}^{1/20}\Gamma_{\rm c, 2.5}^{11/5}n_{\rm amb.,-1}^{\rm ns\ 3/5}\ .
\end{align}

Expressed in terms $\nu_{\rm p, e}^{\rm obs.}$, the key quantities
in the observer's frame (given in the chunk's frame in Appendix \ref{appendix:CSE}) are

\begin{align}
\label{eq:key-equations}
t_{\rm m-WI}^{\rm obs.}&
\simeq 4.3\ {\rm ms}\times \zeta_{\rm m-WI, 5}\times \frac{1}{\nu_{\rm p, e, 9}^{\rm obs.}}\\\nonumber
\mathcal{F}&\simeq 0.7\ {\rm Jy\ ms}\times \frac{\beta_{\rm WI, -1}^{1/2}\zeta_{\rm BI, -1}}{(1+z)^{5/3}f(\Gamma_{\rm c},\theta_{\rm c})^{11/3}d_{\rm L, Gpc}^2}\times \frac{m_{\rm c, 22.3}^{7/12}\Gamma_{\rm c, 2.5}^{11/3}}{{\nu_{\rm p, e, 9}^{\rm obs.}}^{5/3}}\\\nonumber
DM_{\rm c}^{\rm obs.}&\simeq 286.5\ {\rm pc\ cm}^{-3}\times (1+z)^{1/3}f(\Gamma_{\rm c},\theta_{\rm c})^{1/3}
\times \frac{m_{\rm c, 22.3}^{1/3}{\nu_{\rm p, e, 9}^{\rm obs.}}^{4/3}}{\Gamma_{\rm c, 2.5}^{1/3}}\\\nonumber
\Delta t_{\rm BI-WI}^{\rm obs.}&\simeq 32.3\ {\rm days}\times \frac{1}{(1+z)f(\Gamma_{\rm c},\theta_{\rm c})}\times \frac{\Gamma_{\rm c, 2.5}}{{\nu_{\rm p, e, 9}^{\rm obs.}}^2}\ .
\end{align}

The total CSE duration  in our model is  found from $\nu_{\rm CSE, p}(\Delta t_{\rm CSE})=\nu_{\rm p, e}$. We get $\Delta t_{\rm CSE}^{\rm obs.}=t_{\rm m-WI}^{\rm obs.}\times (1/\beta_{\rm WI}^{1/2}-1)$. In the detector's frame, the CSE peak frequency drifts from 
a maximum value of $\nu_{\rm CSE, p, max.}^{\rm obs.}=\nu_{\rm p, e}^{\rm obs.}/\beta_{\rm WI}^{1/2}$
to $\nu_{\rm p, e}^{\rm obs.}$ (see Figure \ref{fig:drifting} and Appendix \ref{appendix:CSE}).  The last equation listed above, $\Delta t_{\rm BI-WI}^{\rm obs.}$, is the radial repetition time in the CSE emission where the chunk experiences multiple BI-WI
episodes as it travels radially away from the QN site (see \S \ref{sec:periodicity}).

 At this point, it is important to mention that the ambient environment in which the QN event takes place, and where the chunks transition into a collisionless state, may differ from the ionized ambient medium (plasma) where the BI-WI and the FRB occur. For instance, a QN chunk can become collisionless within a neutral ambient medium in the galactic disk and subsequently emit a CSE, which is identified as the FRB, when it enters an ionized medium such as the Warm Ionized Medium (WIM) in the disk, the galactic halo, the
  Intergalactic plasma medium (IGpM), and the Intergalactic Medium (IGM) as it moves away from the QN event; see Figure  \ref{fig:WIM-and-Halo}.
 
Consider for example a QN occurring in a neutral medium in the disk where the chunks also become collisionless. 
Once a chunk becomes collisionless, it ceases to expand and retains its plasma frequency for as long as it can emit FRBs while traversing through various ionized media. Apart from the CSE luminosity (and thus fluence), which depends on the 
density of the ionized medium ($n_{\rm ionized}^{\rm ns}$; see Appendix  \ref{appendix:CSE-luminosity}), all other quantities are determined by the electron plasma frequency ($\nu_{\rm p, e}\propto n_{\rm c}^{1/2}\propto n_{\rm amb.}^{3/5}$) and thus by the density of the ambient medium where the QN occurs ($n_{\rm amb.}^{\rm ns}$). 
Hereafter, we continue with the  $n_{\rm ionized}^{\rm ns}= n_{\rm amb.}^{\rm ns}$ scenario. 

\subsection{Radial repetition}
\label{sec:periodicity}

As explained in Appendix \ref{appendix:radial-periodicity},
once the WI saturates and shuts-off, electron Coulomb collisions within a chunk act to dissipate the WI produced magnetic field
on timescales $\Delta t_{\rm BI-WI}^{\rm obs.}$ at which point a new BI-WI episode (followed by CSE)  begins. 
Effectively, $\Delta t_{\rm BI-WI}^{\rm obs.}$ is the magnetic field dissipation timescale. The electron Coulomb collision frequency is proportional to the chunk's density when  it becomes collisionless $\nu_{\rm coll.}\propto n_{\rm c}$ (see Appendix \ref{appendix:radial-periodicity}). This means that the magnetic field dissipation timescale $\Delta t_{\rm BI-WI}^{\rm obs.}\propto \nu_{\rm coll.}^{-1}\propto n_{\rm c}^{-1}\propto  \nu_{\rm p, e}^{-2}$.
   
  The maximum distance travelled by a chunk between CSE episodes is given by Eq. (\ref{eq:distance}) in Appendix \ref{appendix:radial-periodicity} (for $\alpha_{\rm diff.}=1$), as 
  
  \begin{equation}
  \Delta r_{\rm BI-WI}^{\rm ns} \simeq 28.7\ {\rm kpc}\times \frac{1}{\Gamma_{\rm c, 2.5}^{7/5}m_{\rm c, 22.3}^{1/10}n_{\rm amb., -1}^{\rm ns\ 6/5}}\ .
  \end{equation}
 
 Radial repetition occurs if the length of the ambient ionized plasma $L_{\rm amb.}$ exceeds $\Delta r_{\rm BI-WI}^{\rm ns}$ which translates to 
   $n_{\rm amb.}^{\rm ns}> n_{\rm amb. c}^{\rm ns}$ with (see Eq. (\ref{eq:nambc})) 
\begin{equation}
 n_{\rm amb., c}^{\rm ns}\sim 0.31\ {\rm cm}^{-3}\times L_{\rm amb., 10kpc}^{-5/6}\Gamma_{\rm c, 2.5}^{-7/6}m_{\rm c, 22.3}^{-1/12}\ .
 \end{equation}
 This critical density  separates the non-repeating chunks (i.e. radial one-offs) from repeating chunks (i.e. radial repeaters)
  and it is very weakly dependent on the chunk's mass for a fixed $\Gamma_{\rm c}$.

\section{The primary chunk: radial repeaters}
\label{sec:single-chunk}

\subsection{The simulations}
\label{sec:single-chunk-simulations}

Consider a single chunk at a viewing angle $\theta_{\rm c}=0$ (hereafter the primary chunk) as seen by an observer (see Figure \ref{fig:single-chunk}). For a given QN (i.e. for a given $N_{\rm c}$ and $n_{\rm amb.}^{\rm ns}$), different observers
would see primary chunks with the same $\Gamma_{\rm c}$ but different mass $m_{\rm c}$.  
  Different chunks from the same QN may encounter different plasmas as they travel away from the QN site.
 
\begin{figure*} 
 \includegraphics[scale=0.6]{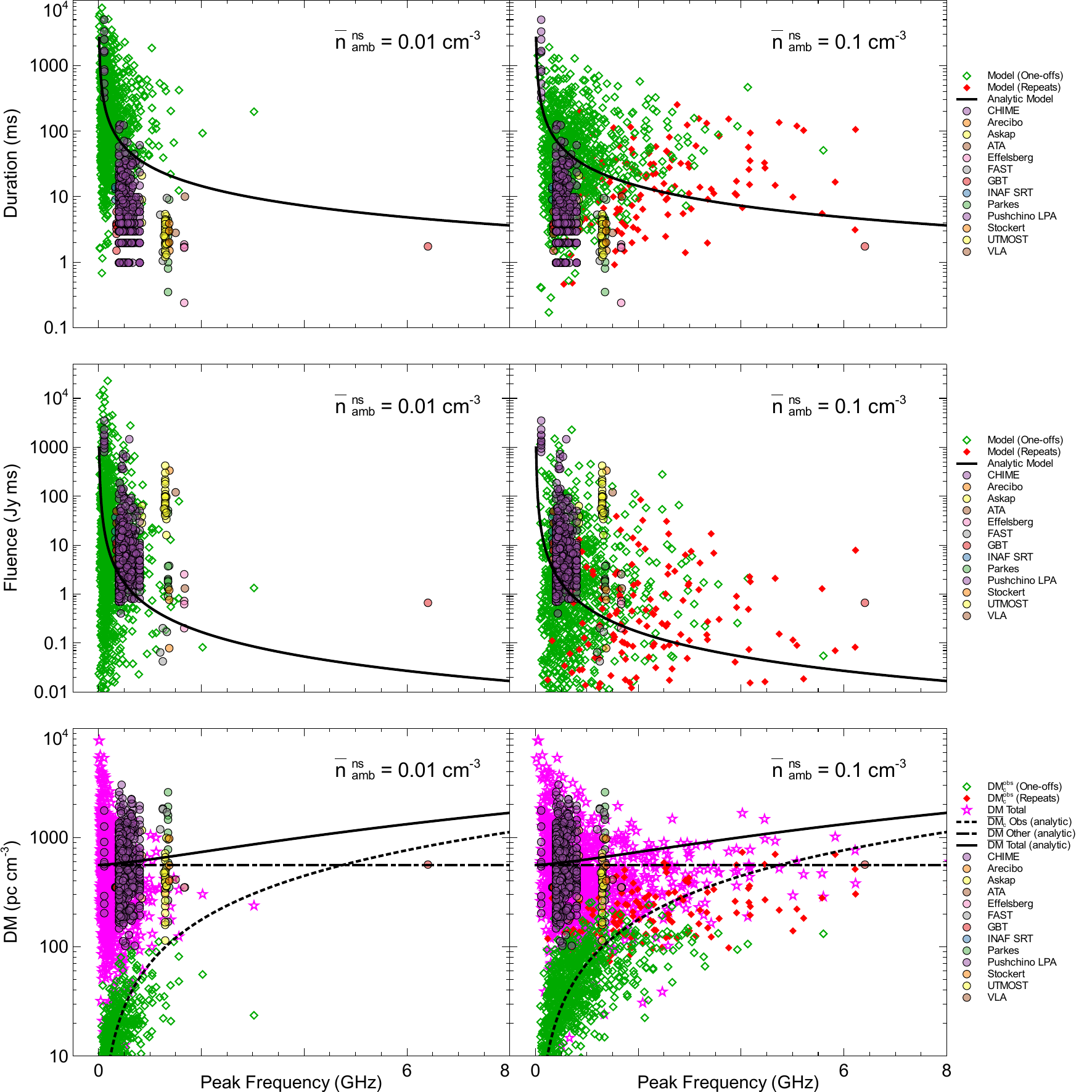} 
 \caption{CSE from the primary chunks from different QNe (open green diamonds are radial one-offs; filled red diamonds are radial repeaters) compared to FRB data (open circles; from FRBSTATS website (\citet{spanakis2021}) and from CHIME Catalog 1 data (\citet{chime2021})).
From top to bottom: duration (ms), fluence (Jy ms) and total DM (pc cm$^{-3}$) versus CSE maximum peak frequency (GHz);  $\nu_{\rm CSE, p, max.}^{\rm obs.}=\nu_{\rm p, e}^{\rm obs.}/\beta_{\rm WI}^{1/2}$. The solid curves are from analytical expressions
given in Eq. (\ref{eq:key-equations}) using mean parameter values. In the DM panels, the dashed line is the DM$_{\rm c}^{\rm obs.}\propto {\nu_{\rm p, e}^{\rm obs.}}^{4/3}$ from Eq. (\ref{eq:key-equations}) while the horizontal dot-dashed line is DM$_{\rm other}={\rm DM}_{\rm host}/(1+z)+{\rm DM}_{\rm MW}+{\rm DM}_{\rm IGM}=532$ pc cm$^{-3}$ using mean values for the parameters;  the total DM, ${\rm DM}_{\rm c}^{\rm obs.}+{\rm DM}_{\rm other}$, is shown by the magenta stars. The left panels are for $\bar{n}_{\rm amb.}^{\rm ns}=0.01$ cm$^{-3}$ and $\bar{n}_{\rm amb.}^{\rm ns}=0.1$ cm$^{-3}$ in the right panels.}
\label{fig:compare-to-HERTA}
\end{figure*}

\begin{figure*} 
 \includegraphics[scale=0.6]{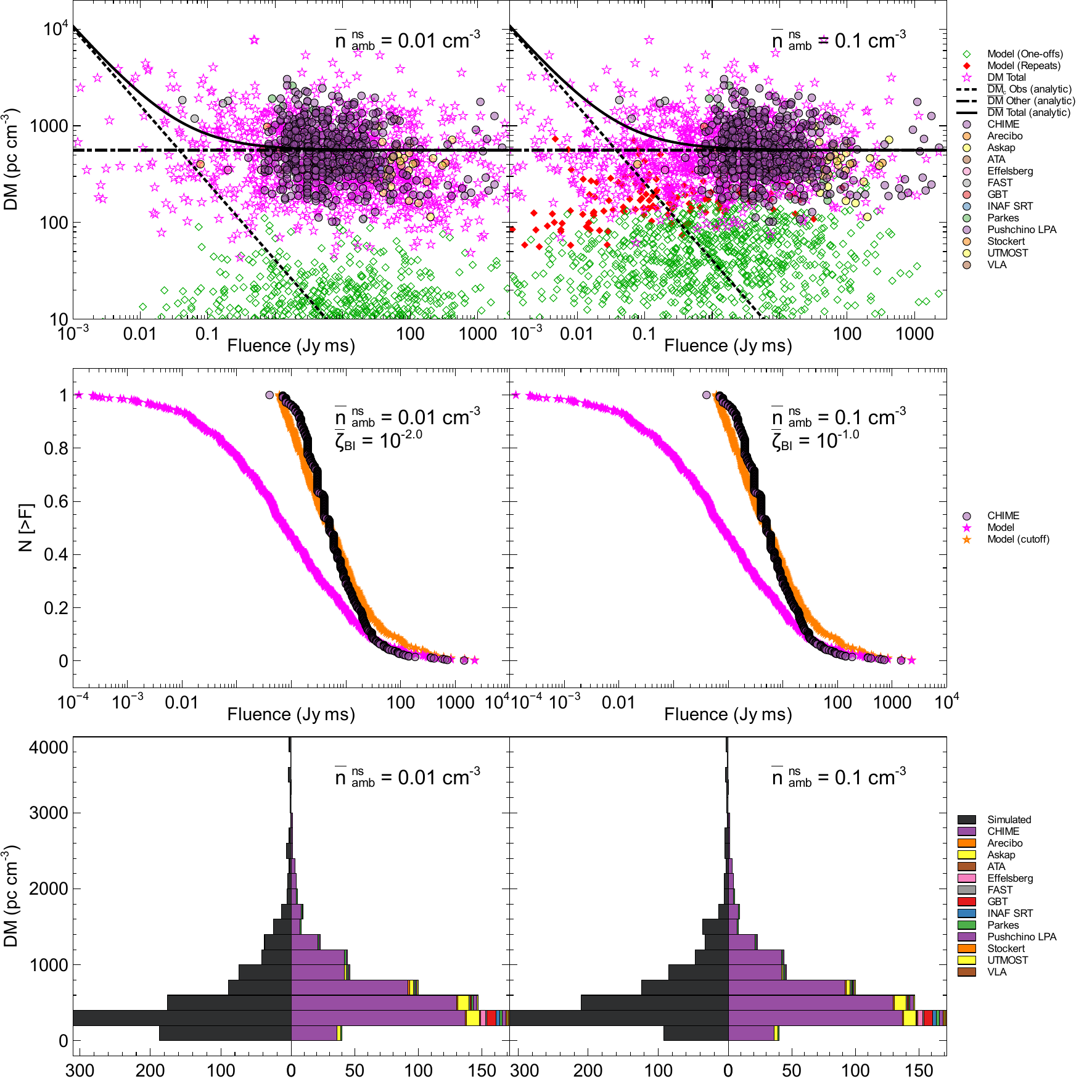} 
 \caption{{\bf Top panels}: Total DM in our model (magenta stars) compared to CHIME FRB (Baseband) data shown as the open circles.
Radial repeaters' DM$_{\rm c}^{\rm obs.}$ is shown by the filled red diamonds while one-offs are the open green diamonds. 
The dashed line is DM$_{\rm c}^{\rm obs.}\propto {\mathcal{F}}^{-4/5}$ from Eq. (\ref{eq:DMc-vs-Fluence}) while the horizontal dot-dashed line is DM$_{\rm other}={\rm DM}_{\rm host}/(1+z)+ {\rm DM}_{\rm MW}+{\rm DM}_{\rm IGM}=532$ pc cm$^{-3}$ from mean values. The analytical total DM is the solid curve. {\bf Middle panels}: The normalized cumulative distribution of burst fluences derived from the simulations (magenta stars) compared to that of CHIME's baseband data (open circles). To compare to data we impose a cutoff on the simulations fluence based on CHIME's fluence detection limit of $\sim 0.6$ Jy ms.  {\bf Bottom panels}:  Histograms of the simulated total DM (left half)  compared to that from FRB data (right half). The left panels are for $\bar{n}_{\rm amb.}^{\rm ns}=0.01$ cm$^{-3}$ and $\bar{n}_{\rm amb.}^{\rm ns}=0.1$ cm$^{-3}$ in the right panels.}
\label{fig:compare-to-HERTA2}
\end{figure*}

 For all of the varied parameters listed in Table \ref{table:fiducial-values} we chose a log-normal distribution with $\sigma (\log(x))=1.0$. We  take a mean fiducial redshift value of $\bar{z}=0.4$ representative of the upper detection distance of FRBs (e.g. \citealt{law2024}). 
We convert the luminosity distance to Mpcs using $d_{\rm L}({\rm Mpc})\simeq 2 z \times (2.4+z)$ (valid for $z<1$; e.g. \citealt{petroff2019} and references therein). The total DM in our model is DM$_{\rm T}= {\rm DM}_{\rm c}^{\rm obs.}+ {\rm DM}_{\rm other}$ where 
 DM$_{\rm other}={\rm DM}_{\rm host}/(1+z)+{\rm DM}_{\rm MW}+{\rm DM}_{\rm IGM}$ is the sum of contributions from 
 the host galaxy (DM$_{\rm host}$), the Milky-Way galaxy (DM$_{\rm MW}$) and the Inter-Galactic medium (DM$_{\rm IGM}$).
We take DM$_{\rm IGM}= z\times 900$ pc cm$^{-3}$ (\citealt{yang2017}) and adopt a normal distribution for the MW and host galaxy
with mean $100$ pc cm$^{-3}$ and $\sigma_{\rm DM}=50$ pc cm$^{-3}$  (e.g. \citealt{taylor1993,yang2016,macquart2020};
see also \citealt{petroff2019}); we ignore possible intervening galaxies close to the l.o.s.. I.e. $\overline{\rm DM}_{\rm other}=532$ pc
cm$^{-3}$ for fiducial mean values.

In Figures \ref{fig:compare-to-HERTA} and \ref{fig:compare-to-HERTA2}, we compare  properties of simulated CSE emission 
in our model to FRB data. The left panels are for $\bar{n}_{\rm amb.}^{\rm ns}=0.01$ cm$^{-3}$ (the mean value) while 
 $\bar{n}_{\rm amb.}^{\rm ns}=0.1$ cm$^{-3}$ in the right panels.
The  data we use (open circles) has been retrieved from FRBSTATS website (\citet{spanakis2021}) and from CHIME Catalog 1 data (\citet{chime2021}) and CHIME Baseband data (\citet{chime2023}).  

Figure \ref{fig:compare-to-HERTA}  is a comparison of our model's duration, fluence and total DM versus the maximum peak frequency ($\nu_{\rm CSE, p, max.}^{\rm obs.}=\nu_{\rm p, e}^{\rm obs.}/\beta_{\rm WI}^{1/2}$) to 
those from FRB data. Detection occurs as long as the detector's upper frequency exceeds the plasma
frequency; $\nu_{\rm max.}^{\rm Det.}>\nu_{\rm p, e}^{\rm obs.}$.
The solid red diamonds show radially repeating primaries (i.e. radial repeaters) while the green diamonds
show the non-repeating primaries (i.e. radial one-offs).  The solid curves are obtained using the parameters
mean values  listed in Table \ref{table:fiducial-values}  and capture only the $\nu_{\rm CSE, p, max.}^{\rm obs.}=\nu_{\rm p, e}^{\rm obs.}/\beta_{\rm WI}^{1/2}$  dependencies as expressed in Eq. (\ref{eq:key-equations}).  The $\bar{n}_{\rm amb.}^{\rm ns}=0.01$ cm$^{-3}$ simulations are all radial one-offs as expected while radial repeaters make up a small percentage of the population 
at $\bar{n}_{\rm amb.}^{\rm ns}=0.1$ cm$^{-3}$. The chunk's DM contribution is more important at higher $n_{\rm amb.}^{\rm ns}$ (i.e. for radial repeaters) and in particular for massive chunks because DM$_{\rm c}^{\rm obs.}\propto m_{\rm c}^{2/5}n_{\rm amb.}^{\rm ns\ 4/5}$.

\begin{figure} 
\centering
 \includegraphics[scale=0.4]{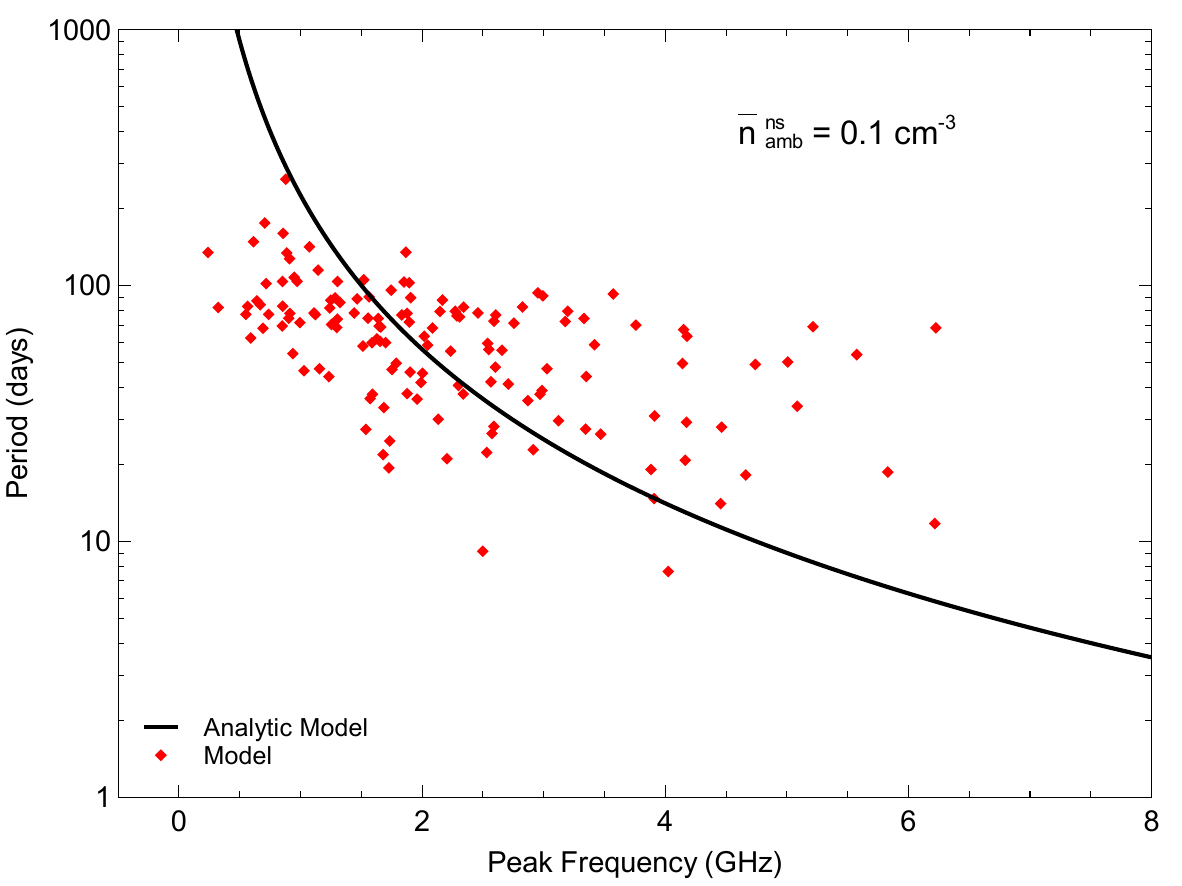} 
 \caption{Radial time interval between BI-WI episodes ($\Delta t_{\rm BI-WI}^{\rm obs.}$; in days) versus peak frequency (GHz).}
\label{fig:period-vs-frequency}
\end{figure}

The top panels in Figure \ref{fig:compare-to-HERTA2} are the simulated DM versus fluence  compared to CHIME Baseband data. The dashed curve is DM$_{\rm c}^{\rm obs.}\propto \mathcal{F}^{-4/5}$ derived from Eq. (\ref{eq:key-equations}) as
\begin{equation}
\label{eq:DMc-vs-Fluence}
DM_{\rm c}^{\rm obs.}\simeq 207\ {\rm pc\ cm}^{-3}\times \mathcal{F}[{\rm Jy\ ms}]^{-4/5}
 \times \frac{\beta_{\rm WI, -1}^{2/5}\zeta_{\rm BI, -1}^{4/5}\Gamma_{\rm c, 2.5}^{13/15}m_{\rm c, 22.3}^{4/5}}{(1+z)f(\Gamma_{\rm c},\theta_{\rm c})^{13/5}d_{\rm L, Gpc}^{8/5}}\ .
\end{equation}
The solid curve is the total DM given as DM$_{\rm T}= {\rm DM}_{\rm c}^{\rm obs.}+{\rm DM}_{\rm other}$ with $\overline{{\rm DM}}_{\rm other}=532$ pc cm$^{-3}$ for mean parameter values shown as the horizontal dot-dashed line in the top panels. 

The middle panels in Figure \ref{fig:compare-to-HERTA2} show the normalized cumulative count distribution of burst fluences ($N(>\mathcal{F})$)  from our simulations (the magenta markers). To compare to CHIME's data we impose a fluence cutoff of 0.6 Jy ms (orange markers) which shows good agreement compared to that of CHIME's Baseband data (\citet{chime2023}). The $\bar{n}_{\rm amb.}^{\rm ns}=0.01$ cm$^{-3}$ 
simulations gave best agreement with CHIME Baseband fluence when using $\zeta_{\rm BI}=10^{-2}$.
Histograms of the simulated total DM  compared to that from FRB data are shown  in the bottom panels of Figure \ref{fig:compare-to-HERTA2} which also shows good agreement.
 
 Figure \ref{fig:period-vs-frequency} shows $\Delta t_{\rm BI-WI}^{\rm obs.}$ for radially repeating chunks (i.e. when $n_{\rm amb.}^{\rm ns} > n_{\rm amb., c}^{\rm ns}$)  versus the CSE peak frequency. With  the $n_{\rm amb.}^{\rm ns} > n_{\rm amb., c}^{\rm ns}$ condition inserted in $\Delta t_{\rm BI-WI}^{\rm obs.}$ in Eq. (\ref{eq:key-equations}) we get an upper limit  
  
\begin{equation}
\Delta t_{\rm BI-WI}^{\rm obs.} < 70\ {\rm days}\times \frac{(1+z)L_{\rm amb., 10kpc}}{\Gamma_{\rm c, 2.5}^{2}}\ .
\end{equation}
For illustrative purposes, the analytical curve in Figure \ref{fig:period-vs-frequency} is extended beyond the limit. The scatter 
is from the $\beta_{\rm WI}$ log-normal distribution and that of redshfit since $\Delta t_{\rm BI-WI}^{\rm obs.} \propto \beta_{\rm WI}/(1+z)\nu_{\rm CSE, p, max.}^2$. In general primary chunks repeat only once with a ``radial period" of tens of days. 
 For a given chunk, $\Delta t_{\rm BI-WI}^{\rm obs.}$ is a constant unless $\alpha_{\rm diff.}$ is allowed to vary between BI-WI episodes which adds to the scatter shown in Figure \ref{fig:period-vs-frequency}.
  
 Overall, and relying solely on the primary chunk, the agreement between our model and FRB data is encouraging. The 
 distribution and  statistical properties of most FRBs we argue is sampling the properties of CSE from primary chunks from different QNe 
 occurring in different environments inside and outside their different host galaxies.

\subsection{The chunk's intrinsic DM}
\label{sec:excess-DM}

Some FRBs show a large excess DM beyond that which is predicted by the Macquart relation 
(i.e. what is expected from cosmological and MW contributions; \citet{macquart2020}). 
This excess could be due to the host galaxy itself, or possible foreground objects such as galaxy halos
 along the line-of-sight (Spitler et al. 2014; Chatterjee et al. 2017; Tendulkar et al. 2017; Hardy et al. 2017; Chittidi et al. 2020).

\begin{figure} 
\centering
\includegraphics[scale=0.4]{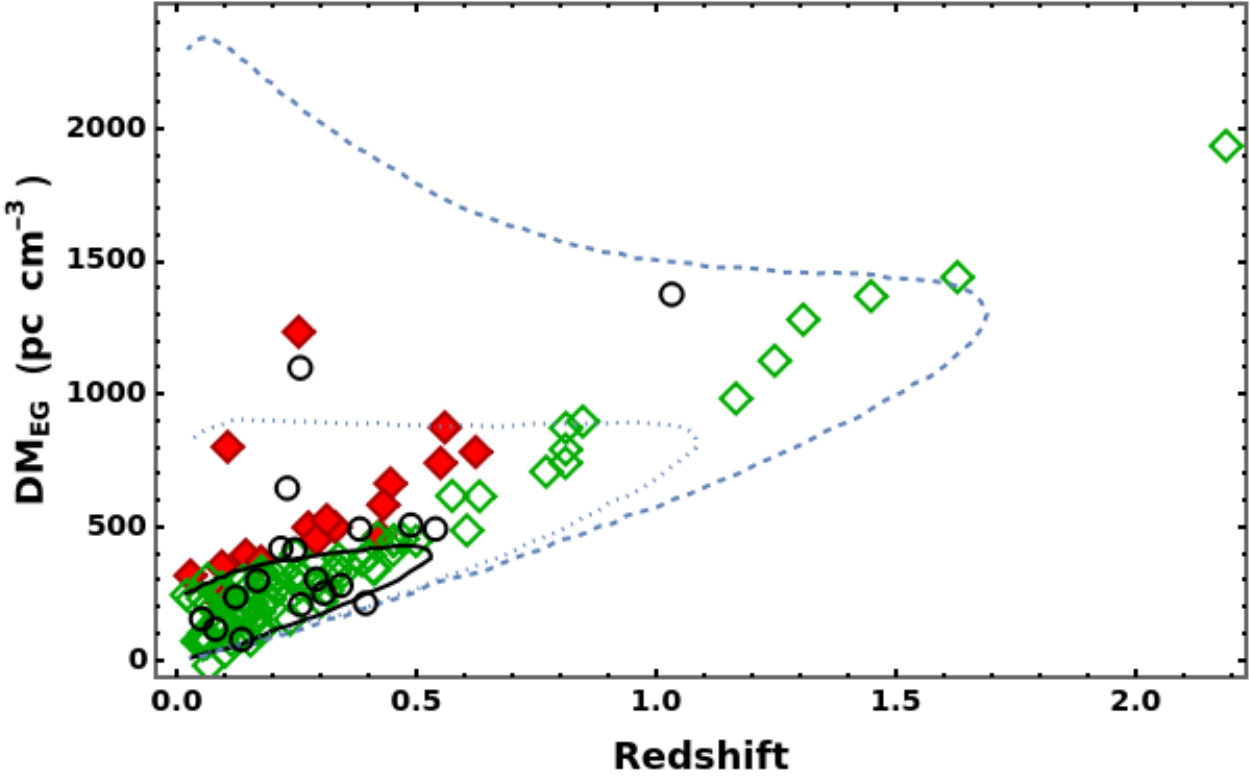}
\caption{The  extragalactic DM (DM$_{\rm EG}={\rm DM}_{\rm IGM}+{\rm DM}_{\rm host}/(1+z)+{\rm DM}_{\rm c}^{\rm obs.}$) versus redshift 
 from the primary chunks. The one-offs are the green diamonds while radially repeating FRBs are shown as red diamonds.
The open circles are data with the curves from top to bottom showing the 99\%, 90\% and 50\%  probability for excess DM due to intervening galaxies along the line-of-sight, respectively (see Figure 3 in \citet{ryder2023}).}
\label{fig:DM-excess-figure}
\end{figure}

Figure \ref{fig:DM-excess-figure} shows the extragalactic DM (DM$_{\rm EG}={\rm DM}_{\rm IGM}+{\rm DM}_{\rm host}/(1+z)+{\rm DM}_{\rm c}^{\rm obs.}$) versus redshift in our model as compared to data (open circles; from \citealt{ryder2023}). 
 The contours in \citet{ryder2023} reflect the ability of ASKAP to detect FRBs, and do not necessarily reflect the intrinsic distribution. A more accurate comparison (a future endeavour) would require placing a cut on fluence in our simulated data.
 
  The contribution of the repeating chunk to the DM is important in some cases and may explain the excess on its own. 
The estimated DM$_{\rm cosmic}$ (observed DM less Galactic and host contributions) for  
 some observed FRBs seems to exceed the average DM$_{\rm cosmic}$ given by the Macquart relation (e.g. \citealt{simha2023}). While such an excess may be  within scatter in IGM DM we argue it may be  due to the QN chunk's intrinsic DM contribution. 
  For example, for $n_{\rm amb.}=1$ cm$^{-3}$, from Eq. (\ref{eq:key-equations}) we get DM$_{\rm c}^{\rm obs.}\sim 446$ pc cm$^{-3}$
  with $\nu_{\rm p, e}^{\rm obs.}\sim 1.4$ GHz. Higher  DM$_{\rm c}^{\rm obs.}$ can be obtained in our model
  by considering higher $\Gamma_{\rm c}$ values.
  
 Evidence for chunk DM contribution can be found in  FRBs where the DM$_{\rm host}$ can be constrained from H$_\alpha$ measurements
(\citealt{cordes2016}) and where DM$_{\rm cosmic}$ far exceeds that derived
from the Macquart relation.

 \subsection{Chunk propagation and DM variation}
  \label{sec:propagation-DM}
  
For this case, we have to consider $n_{\rm ionized}^{\rm ns}\neq n_{\rm amb.}^{\rm ns}$. More specifically, 
$n_{\rm ionized}^{\rm ns}<< n_{\rm amb.}^{\rm ns}$ so that $L_{\rm ionized}>>\Delta r_{\rm BI-WI}^{\rm ns}$ because
radial repeaters which are more likely to emanate from chunks born in disks (i.e. high $n_{\rm amb.}^{\rm ns}$ yielding smaller $ \Delta r_{\rm BI-WI}^{\rm ns}\propto n_{\rm amb.}^{\rm ns\ -6/5}$) but propagating in very low-density ionized environments (i.e. lower $n_{\rm ionized}^{\rm ns}$) such as galactic halos, IGpM  and IGM. The change in DM over $N_{\rm BI-WI}$ BI-WI episodes   is then 
 \begin{align}
 \label{eq:DeltaDM}
 \Delta (DM^{\rm obs.}) &= n_{\rm ionized}^{\rm ns}\times (N_{\rm BI-WI} \Delta r_{\rm BI-WI}^{\rm ns}) \\\nonumber
 &\sim N_{\rm BI-WI}\ {\rm pc\ cm}^{-3}\times \left(\frac{n_{\rm ionized}^{\rm ns}}{10^{-3}\ {\rm cm}^{-3}}\right)\left(\frac{\Delta r_{\rm BI-WI}^{\rm ns}}{{\rm kpc}}\right)\ .
 \end{align}
 We note that $\Delta r_{\rm BI-WI}^{\rm ns}$ is in the sub-kpc if we adopt  $\alpha_{\rm diff.}<1$ (see Eq. (\ref{eq:tBdiff})).
 For example for $\alpha_{\rm diff.}\sim 0.01$, a chunk travelling in a galactic halo can experience $\sim 100$ BI-WI episodes
 before an observer sees noticeable changes in the total DM of the FRBs.
  The maximum variation in the total DM of FRBs from a repeating chunk after crossing the ionized medium (e.g. a galactic halo) 
  is $n_{\rm ionized}^{\rm ns}L_{\rm ionized}\sim (10^{-3}\ {\rm cm}^{-3})\times (10\ {\rm kpc})\sim 10$ pc cm$^{-3}$.
 It is less if the chunk is propagating in the IGpM or in the IGM.

\subsection{The ``sad trombone"}
\label{sec:sadtrombone}

For a given electron bunch of size $\lambda_{\rm b}$, the maximum CSE frequency would drift over time towards the chunk's plasma frequency  as   $\nu_{\rm CSE, p}(t)=c/\lambda_{\rm b}(t)= (\nu_{\rm p, e}/\beta_{\rm WI}^{1/2})\times (1+t/t_{\rm m-WI})^{-\delta_{\rm m-WI}}$; recall that $\delta_{\rm b}=1$.  During the turbulent merging of the WI filaments (and thus of the emitting bunches), while the frequency is drifting over time, intermittent CSE emission occurs. This is  mainly due to  the random orientation of the filaments until they re-arrange themselves in the direction of motion of the chunk and the CSE become observable again. This is illustrated in Figure \ref{fig:drifting}.  
  
\begin{figure*} 
\centering
\includegraphics[scale=0.6]{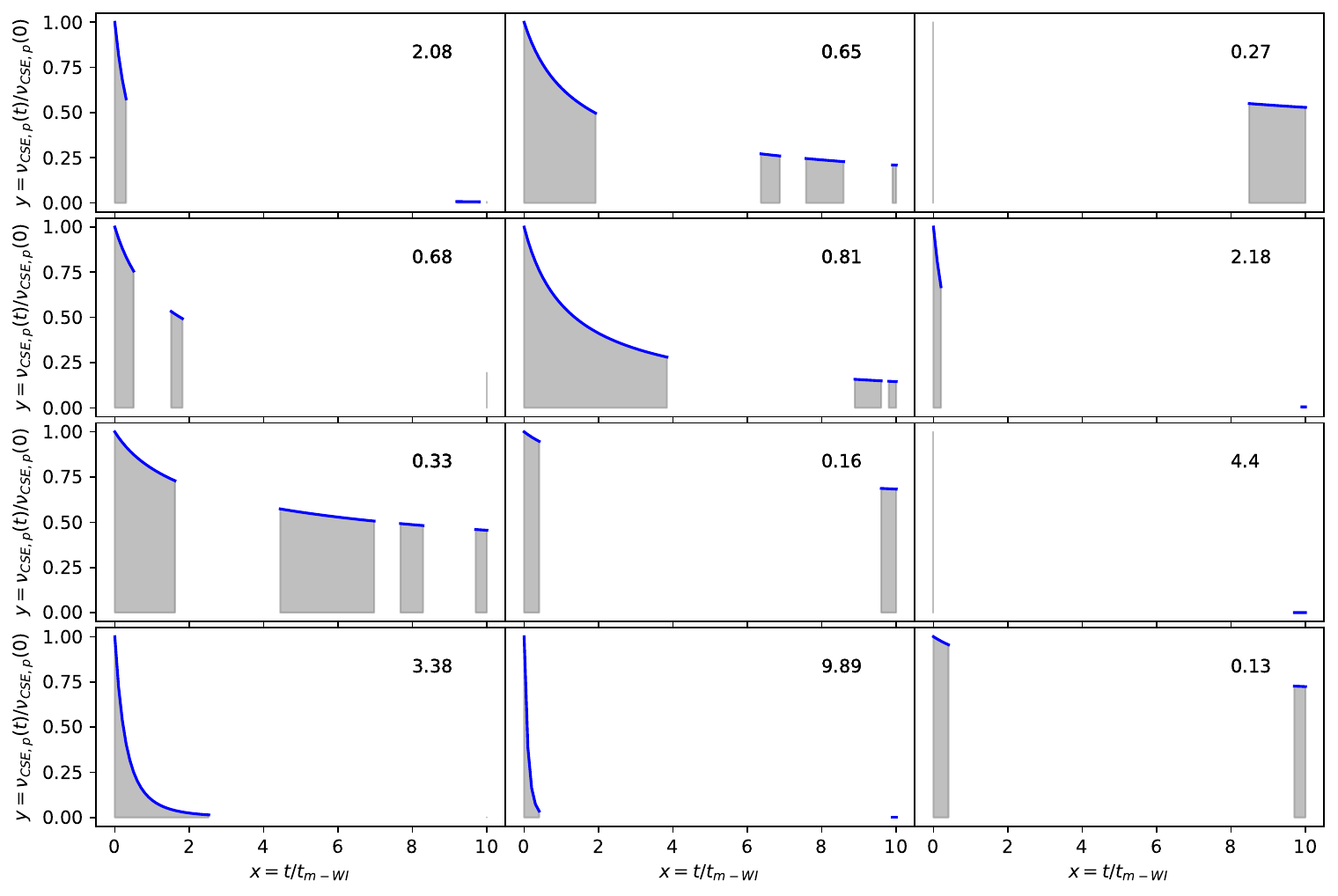}
\caption{CSE frequency (in units of $\nu_{\rm CSE, p, max.}^{\rm obs.}=\nu_{\rm CSE, p}^{\rm obs.}(t^{\rm obs.}=0)$) versus time (in units of $t_{\rm m-WI}^{\rm obs.}$) for a radially repeating primary as it travels in its host galaxy.  The grey shading shows the emission bands while the numbers in each panel gives the value of $\delta_{\rm m-WI}$.}
\label{fig:sadtrombone-primary-flat}
\end{figure*}

\begin{figure*} 
\centering
\includegraphics[scale=0.6]{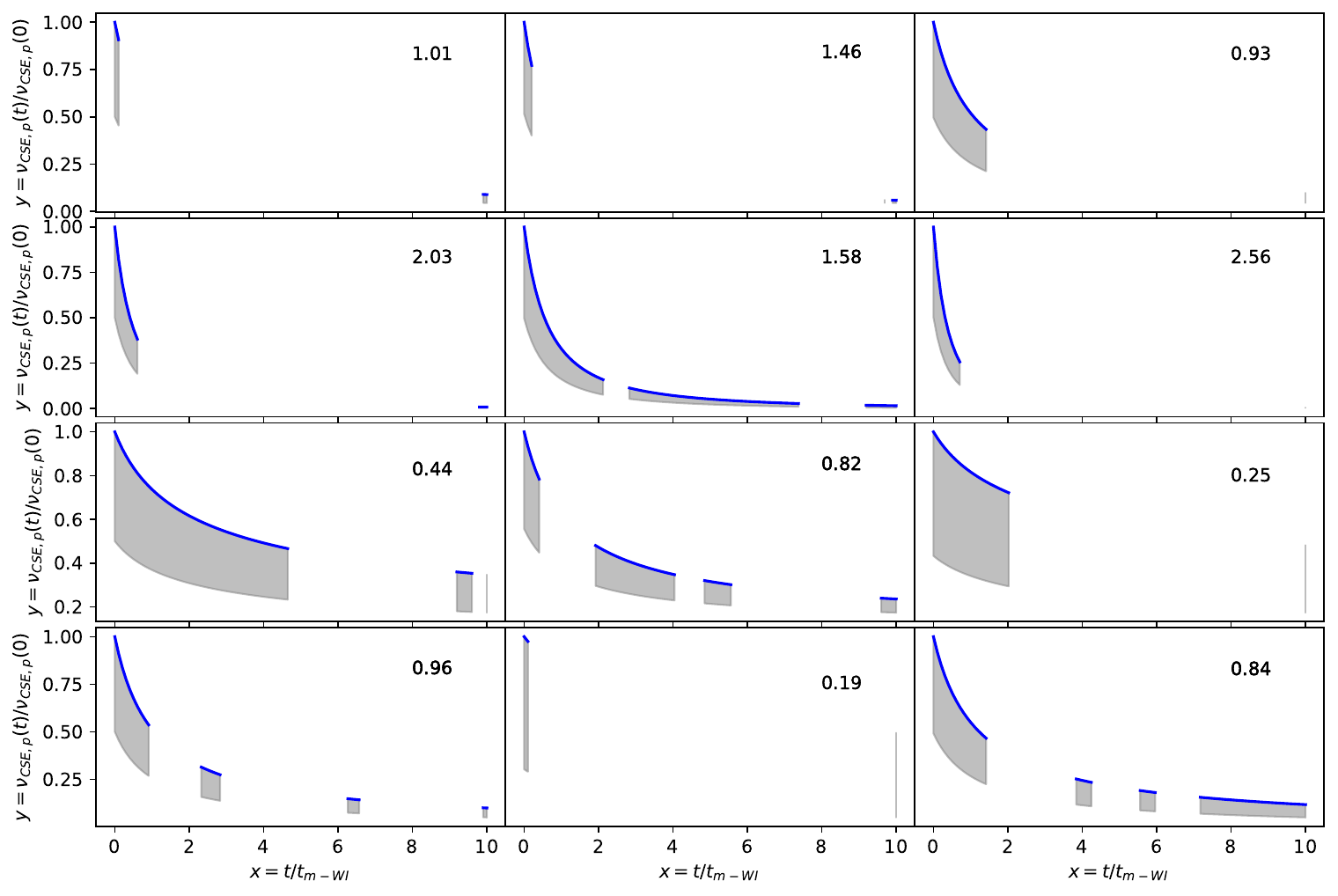}
\caption{Same as in Figure \ref{fig:sadtrombone-primary-flat} but for a power-law spectrum.}
\label{fig:sadtrombone-primary-powerlaw}
\end{figure*}

  Figures  \ref{fig:sadtrombone-primary-flat} and \ref{fig:sadtrombone-primary-powerlaw} show $y=\nu_{\rm CSE}/\nu_{\rm CSE, p, max.}$ versus
 $x=t/t_{\rm m-WI}$ for a primary chunk undergoing multiple BI-WI episodes (the different panels) with filament merging. Between BI-WI episodes, we vary the merging power index $\delta_{\rm m-WI}$ (with the value shown in each panel) using a log-normal distribution with mean value of $1$ and $\sigma{(\log{(\delta_{\rm m-WI}})})=1.0$. For a given episode, the random orientation of the CSE emitting bunches (i.e. on and off CSE emission towards the observer)  is ``mimicked" by  randomly turning the emission on and off for specific intervals of time using a step function; we multiply $y=(1+x)^{-\delta_{\rm m-WI}}$ by a modified step function that accounts for these intervals. The grey shading is purely schematic to highlight emission bands and does not represent any physical quantity.
  
In the spectrum, the minimum CSE frequency is the chunk's plasma frequency while the peak frequency scales as $\nu_{\rm CSE, p, max.}=\nu_{\rm p, e}/\beta_{\rm WI}^{1/2}$. Figure \ref{fig:sadtrombone-primary-flat} illustrate  the case of a flat spectrum while 
 Figure \ref{fig:sadtrombone-primary-powerlaw} is for the case of a power-law spectrum. The spectral nature influences our outcomes, as illustrated in Figure \ref{fig:drifting}. In a power-law spectrum (where the emission weakens at lower frequency), radiation might fade before reaching the plasma frequency.
 
  Figure \ref{fig:sadtrombone-primary-flat} shows an example of an intermittent CSE emission from a radially repeating primary chunk which reproduces the observed ``sad trombone" effect in FRBs.  It represents the scenario where the plasma frequency is below the detector's lower limit. Figure \ref{fig:sadtrombone-primary-powerlaw}  depicts the case when the plasma frequency falls within the detector's band. E.g. for a power-law spectrum or when the plasma frequency is below detector's sensitivity cutoff.
 
  From a BI-WI episode to another all three scenarios  of CSE (no emission, a single sub-pulse and many sub-pulses per episode)
 are reproduced randomly.  Sub-pulses much shorter than $t_{\rm m-WI}$ 
 (i.e. in the sub-microsecond regime in the observer's frame where $t_{\rm m-WI}^{\rm obs.}\sim $ ms)
  can be reproduced while in some cases the drifting appears smooth with no interruption in the CSE.
    In our model, pulses as short as the  bunch's cooling timescale are possible.  A bunch 
  cools on timescales $N_{\rm e, b}\gamma_{\rm e}m_{\rm e}c^2/L_{\rm CSE}\sim 10^{-13}$ s 
  where $N_{\rm e, b}$ is the number of electrons per bunch given in Appendix \ref{appendix:Neb-and-NbT}
  and $\gamma_{\rm e}\sim 10$ the electron's Lorentz factor.
 
   Figures  \ref{fig:sadtrombone-primary-flat} and \ref{fig:sadtrombone-primary-powerlaw} can also be interpreted as CSE emission from different one-off primary chunks from different QNe. A parameter survey, using primary chunks only, reproduces 
   most of the sub-pulse behavior and properties observed in repeating FRBs.

 The sad trombone effect will be more prominent in repeaters because of the
 higher plasma frequency ($\nu_{\rm p, e}\propto n_{\rm c}^{1/2}\propto n_{\rm amb.}^{\rm ns}$ with $n_{\rm amb.}^{\rm ns}> n_{\rm amb., c}^{\rm ns}$) which means that a detector will capture the longer end-tail of the pulse (see bottom panel in Figure \ref{fig:drifting} and related discussion).
  Non-repeaters will be caught by a detector earlier and would appear narrower. Because of the wider
  pulse, on average, the sad trombone effect is more likely to be observed in repeaters. This also explains why sub-pulses are in-band
   as the emission gets randomly cut-off during merging.  For individual bursts both repeaters and non-repeaters show the sad trombone, but statistically, the repeaters should show it more often. However, the sad trombone  may not be unique to repeaters.

For completeness, and from $\nu_{\rm CSE, p}^{\rm obs.}=\nu_{\rm CSE, p, max.}^{\rm obs.}\times (1+t^{\rm obs.}/t_{\rm m-WI}^{\rm obs.})^{-\delta_{\rm m-WI}}$, we derive a relationship between the drifting slope and the  CSE frequency
\begin{align}
\frac{d \nu_{\rm CSE}^{\rm obs.}}{d t^{\rm obs.}} &= -\frac{{\nu_{\rm CSE}^{\rm obs.}}^2}{\nu_{\rm CSE, p, max.}t_{\rm m-WI}^{\rm obs.}}\\\nonumber
&\simeq -\left(\frac{7.3\times 10^{-2}\beta_{\rm WI, -1}^{1/2}}{\zeta_{\rm m-WI, 5}}  {\rm GHz^{-1} ms^{-1}}\right)\times {\nu_{\rm CSE}^{\rm obs.}}^2\ .
\end{align}

The above slope should not be confused with the one measured within sub-bursts from different sources (e.g. \citealt{brown2024}).
The above drift is for a given train of sub-bursts within a given source as illustrated in Figure \ref{fig:drifting} and
simulated in Figures \ref{fig:sadtrombone-primary-flat} and \ref{fig:sadtrombone-primary-powerlaw}. 
It should be verifiable with FRB data which we intend to follow up. 

In our model, the upper frequency drifts towards the chunk's plasma frequency which is constant for a given chunk. This is different from the sad-trombone effect in FRBs where  both upper and lower frequencies of bandwidth-limited emission drifting downwards at a roughly constant $d\nu/dt$. The spectrum (flat vs. power-law) may give us what is observed in FRBs as illustrated in Figure \ref{fig:drifting}. It is also possible that a missed physical mechanism/plasma effect may cause the lower frequency to drift down which we plan to explore in the future.

\section{Multiple chunks: Angular repeaters}
\label{sec:multiple-chunks}

We now consider multiple chunks  from a given QN as depicted in Figure \ref{fig:multiple-chunks}. This scenario has been studied in \citet{ouyed2021}. However, in this analysis, we specifically examine the influence of the chunk's intrinsic DM on DM$_{\rm T}$ and investigate the conditions under which ``angular + radial" repetition is feasible. 

   The chunk closest to the line-of-sight is the primary chunk (denoted by the subscript ``P"; the one discussed in the
previous section). The secondary and tertiary chunks, at higher viewing angle, denoted by ``S" and ``T".  
Each chunk from a given QN obeys the relationships given in Eq. (\ref{eq:key-equations}).
A secondary and tertiary chunk differs from the primary by the mass (which is random) and the viewing angle. Because $\Gamma_{\rm c}$ is fixed, and based on average viewing angles, we have  $f(\Gamma_{\rm c}, \theta_{\rm P})\sim 1+ 0.18\times \Gamma_{\rm c, 2.5}^2/N_{\rm c, 6}$ for P,  $f(\Gamma_{\rm c}, \theta_{\rm S})\simeq  1+ 0.97\times \Gamma_{\rm c, 2.5}^2/N_{\rm c, 6}$ for S and 
   $f(\Gamma_{\rm c}, \theta_{\rm T})\simeq  1+ 6.67\times \Gamma_{\rm c, 2.5}^2/N_{\rm c, 6}$ for T   (see Appendix \ref{appendix:angular-periodicity}). 
 
  It is instructive to rewrite the key equations in their general forms where the chunk's plasma frequency is expanded (see Eq. (\ref{eq:plasma-frequency})).  I.e.

 \begin{align}
\label{eq:key-equations-2}
t_{\rm m-WI}^{\rm obs.}&
\simeq 12.3\ {\rm ms}\times \zeta_{\rm m-WI, 5}\times \frac{(1+z)f(\Gamma_{\rm c},\theta_{\rm c})}{m_{\rm c, 22.3}^{1/20}\Gamma_{\rm c, 2.5}^{11/5}n_{\rm amb.,-1}^{\rm ns\ 3/5}}\\\nonumber
\mathcal{F}&\simeq 3.8\ {\rm Jy\ ms}\times \frac{\beta_{\rm WI,-1}^{1/2}\zeta_{\rm BI, -1}}{(1+z)f(\Gamma_{\rm c},\theta_{\rm c})^2 d_{\rm L, Gpc}^2}\times \frac{m_{\rm c, 22.3}^{1/2}}{n_{\rm amb., -1}^{\rm ns}}\\\nonumber
DM_{\rm c}^{\rm obs.}&\simeq 71.2\ {\rm pc\ cm}^{-3}\times  \frac{m_{\rm c, 22.3}^{2/5} \Gamma_{\rm c, 2.5}^{13/5}n_{\rm amb., -1}^{\rm ns\ 4/5}}{(1+z)f(\Gamma_{\rm c},\theta_{\rm c})}\\\nonumber
\Delta t_{\rm BI-WI}^{\rm obs.}&\simeq 267.8\ {\rm days}\times  \frac{(1+z)f(\Gamma_{\rm c},\theta_{\rm c})}{m_{\rm c, 22.3}^{1/10}\Gamma_{\rm c, 2.5}^{17/5}n_{\rm amb.,-1}^{\rm ns\ 6/5}}\\\nonumber
 \Delta t_{\rm ang.}^{\rm obs.}&\simeq 3.0\ {\rm hrs}\times (1+z)\times \frac{m_{\rm c, 22.5}^{1/5}}{N_{\rm c, 6}\Gamma_{\rm c, 2.5}^{1/5}n_{\rm amb, -1}^{\rm ns\ 3/5}}\ .
\end{align}
 We added the last equation which is the average angular time separation, $\Delta t_{\rm ang.}^{\rm obs.}$, between chunks from the same QN which is independent of $f(\Gamma_{\rm c},\theta_{\rm c})$  (see Appendix \ref{appendix:angular-periodicity}).
Except for $\mathcal{F}\propto m_{\rm c}^{1/2}$ and DM$_{\rm c}^{\rm obs.}\propto m_{\rm c}^{2/5}$, all other quantities are weakly dependent
on the chunk's mass; $n_{\rm amb.}^{\rm ns}$ is fixed for a given QN. The two quantities which are prone to the most scatter are $t_{\rm m-WI}^{\rm obs.}$ (and thus the FRB duration) and $\mathcal{F}$ due to their dependency on the stochastic plasma parameters.

For  $n_{\rm amb.}^{\rm ns}< n_{\rm amb., c}^{\rm ns}$ and $N_{\rm c}\sim 10^6$, we get the regime
 of angular repetition only where the chunks DM is negligible because $f(\Gamma_{\rm c},\theta_{\rm c})>>1$ (${\rm DM}_{\rm c}^{\rm obs.}\propto f(\Gamma_{\rm c},\theta_{\rm c})^{-1}$). In this case, all emitting chunks from a given QN would be 
  associated with the same total DM given by DM$_{\rm T}={\rm DM}_{\rm other}>>{\rm DM}_{\rm c}^{\rm obs.}$.
  Also, there is no change in DM$_{\rm T}$ due to propagation because the chunks emit once.  However, angular repetition is at best quasi-periodic and  even irregular because the chunks are not precisely equally spaced and the variation in $f(\Gamma_{\rm c},\theta_{\rm c})$ between  chunks is somewhat variable (see \citet{ouyed2021}). Below we focus on the $n_{\rm amb.}^{\rm ns}> n_{\rm amb., c}^{\rm ns}$  
 scenario where all chunks from a given QN repeat radially. These would correspond to Disk-Born QNe, associated with NSs with small $v_{\rm kick}$, which can subsequently yield Disk-Bursting and Halo-Bursting chunks. However, to have multiple CSE episodes per chunk, the ionized medium is such that $L_{\rm ionized}>> \Delta r_{\rm BI-WI}^{\rm ns}$ which suggest  Disk-Born-Halo-Bursting chunks as the most likely scenario for yielding ``angular$+$radial" repetition. We argue in \S \ref{sec:16p3days} that FRB 180916.J0158$+$65 may be one such candidate.

\subsection{Simulations}
\label{sec:multiple-chunks-simulations}

Shown and discussed here are two  types of simulations based on Eq. (\ref{eq:key-equations-2}).
We consider the same QN (fixed $N_{\rm c}$ and $\sigma (\log{(m_{\rm c})})$) in different ambient environments (i.e. varying $n_{\rm amb.}^{\rm ns}$) while the second set of simulations consider different QNe (different $N_{\rm c}$ and $\sigma (\log{(m_{\rm c})})$) in the same ambient environment at fixed $n_{\rm amb.}^{\rm ns}$.
For simplicity we fix the redshift at $z=0.4$ so that ${\rm DM}_{\rm IGM}\sim 900 z\sim 360$ pc cm$^{-3}$ and one is free to randomly
choose ${\rm DM}_{\rm host}$ and ${\rm DM}_{\rm MW}$ to assign a value for ${\rm DM}_{\rm other}$. The plasma parameters are kept stochastic with mean values and deviation given in Table \ref{table:fiducial-values}. Unless clearly differentiated due to viewing angle effect (i.e. via the $f(\Gamma_{\rm c},\theta_{\rm c})$ dependencies in Eq. (\ref{eq:key-equations-2})), the peak frequency, the duration and the fluence are prone to the most scatter  because of their dependencies on stochastic plasma parameters (see Eq. (\ref{eq:key-equations-2})). These would appear scrambled in our plots. 

For clarity, and for the simulations discussed here, we only plot properties of FRBs
from the primary chunk (P; $\bar{\theta}_{\rm P}\sim 4/3N_{\rm c}^{1/2}$), a single secondary chunk (S; $\bar{\theta}_{\rm S}\sim 2.4\bar{\theta}_{\rm P}$) and tertiary chunk (T; $\bar{\theta}_{\rm T}\sim 6\bar{\theta}_{\rm P}$); see Appendix \ref{appendix:angular-periodicity} for ejecta geometry and chunk's mean viewing angles. Furthermore, only  three BI-WI episodes are shown for each simulated case. For each chunk, 
we assign a radial time $i\times\Delta t_{\rm BI-WI}^{\rm obs.}$ where $i$ is index of the BI-WI bursting episode.  

In the NS frame, the distance travelled by any chunk between BI-WI episodes, $\Delta r_{\rm BI-WI}^{\rm ns}$, is independent of the viewing angle  and it is the same for all chunks for a given QN (see Eq.(\ref{eq:distance})). It is also very weakly dependent on mass ($\propto m_{\rm c}^{-1/10}$). This means that if the primary  is a radial repeater  then all peripheral (here S and T) chunks will also be radial repeaters. 

\begin{figure*} 
\centering
\includegraphics[scale=0.7]{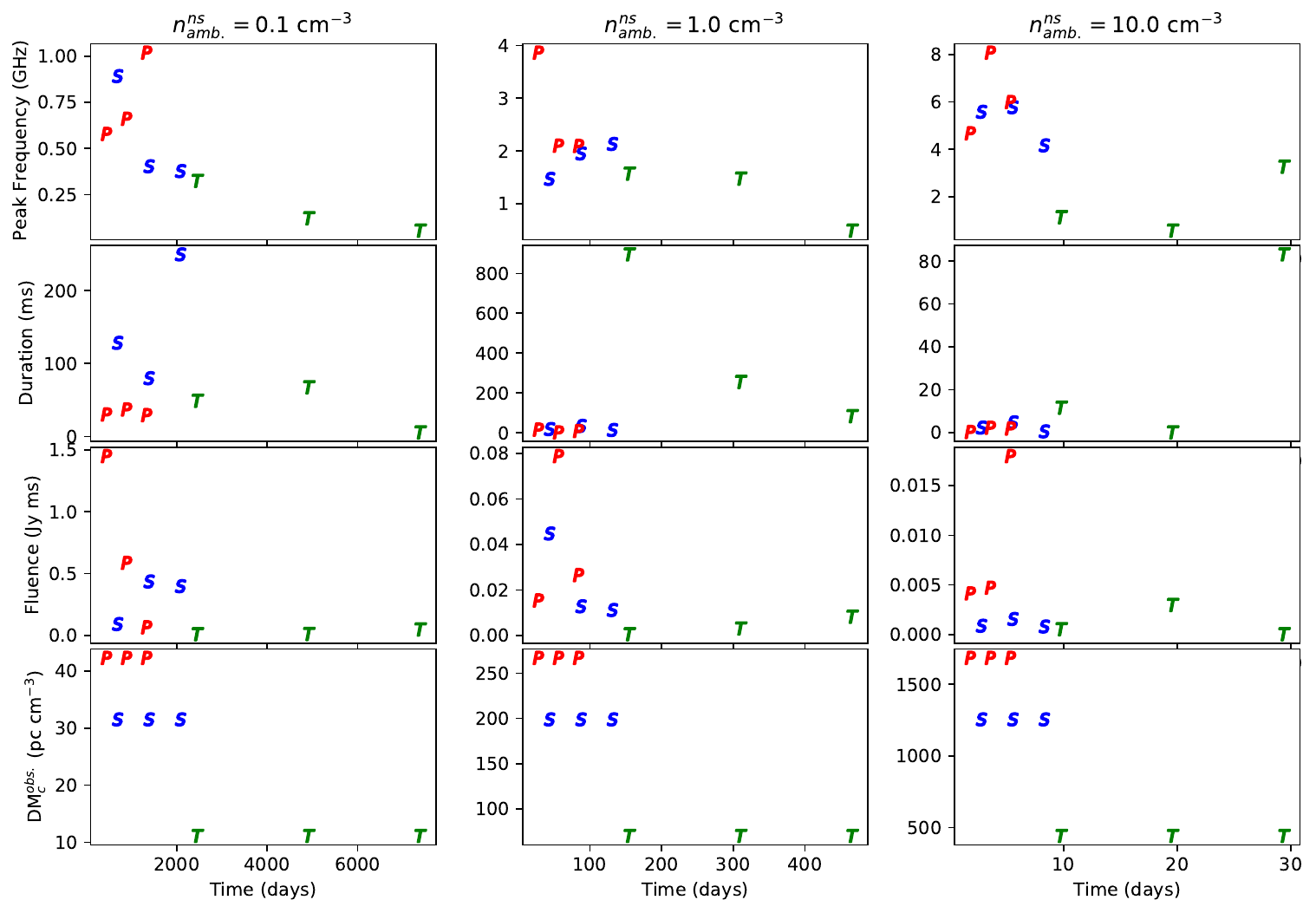}
\caption{FRBs from the primary (P), secondary (S) and tertiary (T) chunks for $n_{\rm amb.}^{\rm ns}=(0.1,1.0,10.0)$ cm$^{-3}$  (from left to right). Three BI-WI episodes are shown for each column.}
\label{fig:Periodicity-fiducial-values-10^6}
\end{figure*}

Figure \ref{fig:Periodicity-fiducial-values-10^6} shows radially repeating FRBs from the P, S and T chunks from a QN at
$z=0.4$ and for $N_{\rm c}=10^6$. The columns from left to right are
for a fixed value of $n_{\rm amb.}^{\rm ns} ({\rm cm}^{-3})=(0.1,1.0,10.0)$. 
From top to bottom are, the frequency, duration, fluence and chunks'  
DM versus time (; i.e. $i\times \Delta t_{\rm BI-WI}^{\rm obs.}$ in days). 
Here, the mass of the P, S and T chunks, once randomly drawn from the log-normal
   distribution with $N_{\rm c}=10^6$ and $\sigma{(\log{(m_{\rm c})})}=1.0$, are kept the same as $n_{\rm amb.}^{\rm ns}$ varies.
    I.e. the change in ${\rm DM}_{\rm c}^{\rm obs.}$ from one column to another is mainly due to changes in $n_{\rm amb.}^{\rm ns}$.

Because $N_{\rm c}=10^6$,  we have $f(\theta_{\rm P})\sim 1, f(\theta_{\rm S})\sim 2, f(\theta_{\rm T})\sim 7$ which helps in clearly distinguishing 
 the P, S and T chunks via the viewing angle effect. This is the case even in the
 frequency, duration and fluence panels despite the large scattering induced by the stochastic plasma parameters.
 The mass distribution has much less of an effect since 
  $\nu_{\rm p, e}\propto m_{\rm c}^{1/20}$ and DM$_{\rm c}^{\rm obs.}\propto m_{\rm c}^{2/5}$. 
 
 Their ``radial period" will differ because $\Delta t_{\rm BI-WI}^{\rm obs.}\propto f(\Gamma_{\rm c},\theta_{\rm c})$. They are less distinguishable  in the duration and fluence panels due to their dependence on the stochastic plasma physics parameters. 
 The viewing angle effect is noticeable even in the scrambled frequency, duration and fluence panels where on average 
  the higher (smaller) frequency (fluence),  FRBs arrive first.
  
   The time interval between BI-WI episodes (i.e. $\Delta t_{\rm BI-WI}^{\rm obs.}$) for different chunks is also more apparent  as expected 
   because $\Delta t_{\rm BI-WI}^{\rm obs.}(\theta_{\rm c})\propto f(\Gamma_{\rm c},\theta_{\rm c})$.
   The angular repetition  timescales (Eq. (\ref{eq:Pcang})), from the left to the right column are $\Delta t_{\rm ang.}^{\rm obs.}\propto n_{\rm amb.}^{\rm ns\ -3/5}\sim 3$ hours, 0.76 hours and
0.2 hours, respectively.
 
  For $n_{\rm amb.}^{\rm ns}=0.1$ cm$^{-3}$ the variation in chunks DM is  narrower but it is still noticeable. 
 As $n_{\rm amb.}^{\rm ns}$ increases, DM$_{\rm c}^{\rm obs.}\propto n_{\rm amb.}^{\rm ns\ 4/5}$ becomes more important and 
 varies widely between chunks. This translates to a wide variation in chunks total DM. In this scenario, 
all chunks repeat radially and despite belonging to the same QN (i.e. the same FRB source) they could
be interpreted as being from different sources and located at different distances.

\begin{figure*} 
\centering
\includegraphics[scale=0.7]{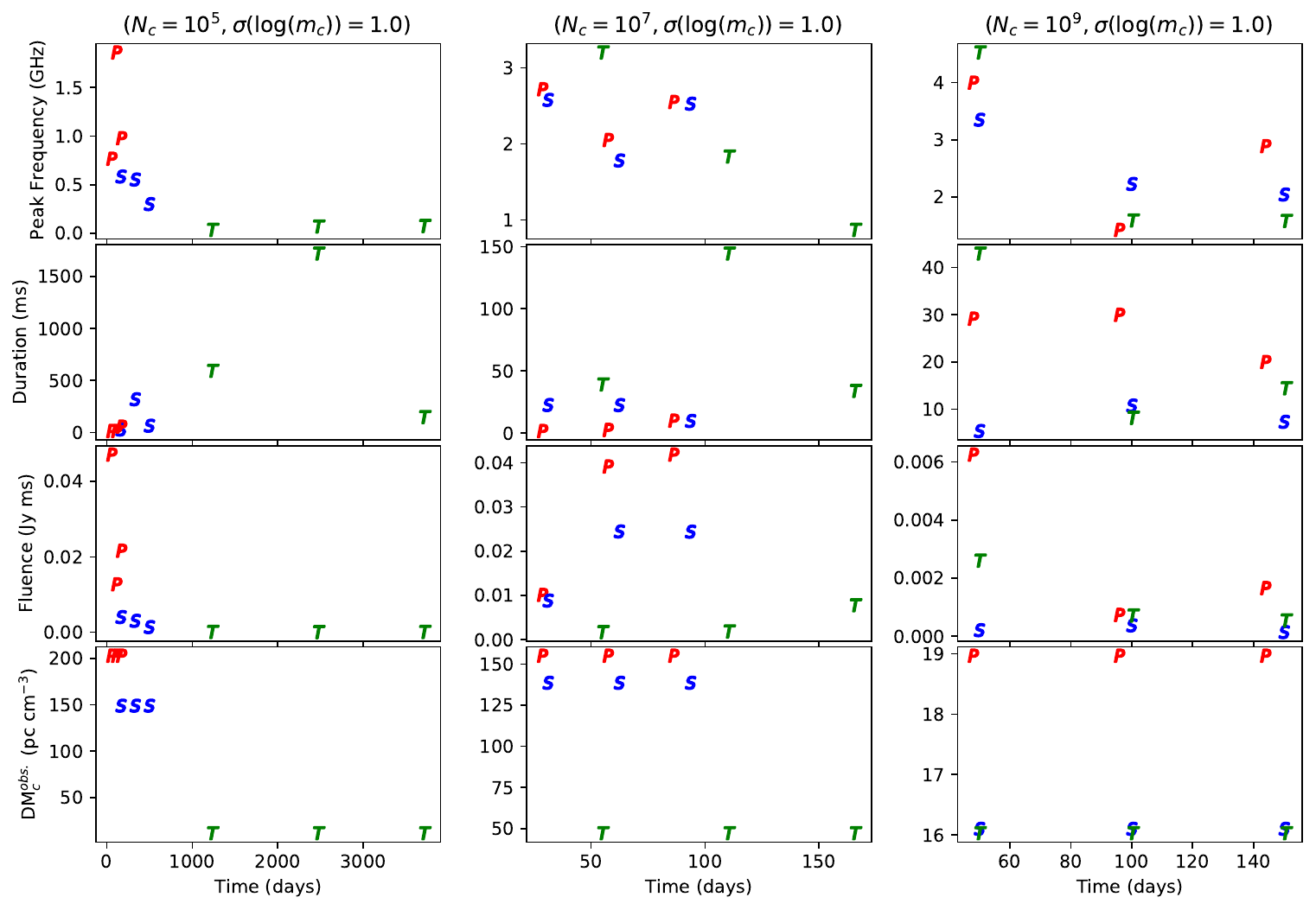}
\caption{FRBs from the primary, secondary and tertiary chunks for $n_{\rm amb.}^{\rm ns}=1$ cm$^{-3}$ and $N_{\rm c}=(10^5,10^7,10^{9})$  (from left to right).  Three BI-WI episodes are shown for each column.}
\label{fig:Periodicity-fiducial-values-10^10}
\end{figure*}

Based on FRB observations, such a large variation in intrinsic chunk DM rules out the $n_{\rm amb.}^{\rm ns}> 0.1$ cm$^{-3}$ scenario. However,  higher $n_{\rm amb.}^{\rm ns}$ values are allowed for high $N_{\rm c}$ when viewing angle effects are negligible.
 Recall that $f(\Gamma_{\rm c},\bar{\theta}_{\rm T})\simeq  1+6.67 \Gamma_{\rm c, 2.5}^2/N_{\rm c, 6}$ (see Appendix \ref{appendix:angular-periodicity}). I.e. the viewing angle effect can be ignored up to the tertiary chunks if $N_{\rm c}> 10^8\Gamma_{\rm c, 2.5}^2$ since
 $f(\Gamma_{\rm c},\theta_{\rm c})\sim 1$. In this case,  ${\rm DM}_{\rm c}^{\rm obs.}\propto m_{\rm c}^{2/5}/f(\Gamma_{\rm c},\theta_{\rm c})\propto m_{\rm c}^{2/5}$  and the intrinsic DM is the same for S, P and T chunks with variations in total DM between chunks due only to propagation in the ambient ionized medium (see Eq. (\ref{eq:DeltaDM})).
 
 Figure \ref{fig:Periodicity-fiducial-values-10^10} shows radial repetition from the P, S and T chunks from a QN at $z=0.4$ and for 
 $n_{\rm amb.}^{\rm ns}=1.0$ cm$^{-3}$. The columns from left to right are for $N_{\rm c} =(10^5,10^7,10^9)$. 
From top to bottom are, the frequency, duration, fluence and chunks'  
DM versus time ($i\times\Delta t_{\rm BI-WI}^{\rm obs.}$  in days).  Here, the mass of the P, S and T chunks, changes from one column to another due to changes in  $N_{\rm c}$.
I.e. the change in ${\rm DM}_{\rm c}^{\rm obs.}$ from one column to another is due to changes in $m_{\rm c}$.

The time interval between BI-WI (i.e. $\Delta t_{\rm BI-WI}^{\rm obs.}$) between different chunks is more apparent at low $N_{\rm c}$ where $f(\Gamma_{\rm c},\theta_{\rm c})>>1$ and viewing effect are most noticeable. 
As $N_{\rm c}$ increases and the $\theta_{\rm c}$  dependence is removed (i.e. $f(\Gamma_{\rm c},\theta_{\rm c})\sim 1$), all chunks start to behave as primaries. One consequence is the  narrowing of  the emission window (in the time axis) despite the vertical scatter due to stochastic plasma parameters. The angular period $\Delta t_{\rm ang.}^{\rm obs.}\propto N_{\rm c}^{-6/5}$ varies from $\sim 12$ hours when
 $N_{\rm c}=10^5$ to $\sim 0.76$ s when $N_{\rm c}=10^9$.  
 
 Despite the weak dependency of ${\rm DM}_{\rm c}^{\rm obs.}$ on mass for  $N_{\rm c}> 10^8\Gamma_{\rm c, 2.5}^2$  (i.e. ${\rm DM}_{\rm c}^{\rm obs.}\propto m_{\rm c}^{2/5}$), the mass distribution yields a variation in ${\rm DM}_{\rm c}^{\rm obs.}$ between chunks from the
 same QN as can be seen even for $N_{\rm c}=10^9$ in the right-most column in Figure \ref{fig:Periodicity-fiducial-values-10^10}. 
 As we show next, high $N_{\rm c}$ with $\log{(\sigma_{\rm m})} < 0.1$ yields a scenario of QN chunks having roughly the same 
 $\nu_{\rm p, e}$ and thus similar ${\rm DM}_{\rm c}^{\rm obs.}$. During multiple BI-WI episodes, FRBs from the QN chunks would only be distinguished due to variations in the duration,  fluence and frequency induced by variations in plasma parameters.

\begin{table*} 
\caption{A 16.3 day periodic activity window in our model (reproducing FRB 180916.J0158$+$65; see Figure \ref{fig:Periodicity-fiducial-values-10^10-16.3days})}
\begin{center}
\begin{tabular}{|c|c|c|c|c|c|c|}\hline
 $N_{\rm c}$  & $n_{\rm amb.}^{\rm ns}$ (cm$^{-3})$ & $\alpha_{\rm diff.}$ &  z & DM$_{\rm MW}$ (pc cm$^{-3}$) & DM$_{\rm host}$ (pc cm$^{-3}$) & DM$_{\rm IGM}$ (pc cm$^{-3}$)\\\hline 
 $10^{8}$  & 1.0&  0.6& 0.0337 & 150 & 100 & $z\times 900\sim 30$ pc cm$^{-3}$\\\hline
\end{tabular}\\
\end{center}
\label{Table:16p3FRB}
\end{table*}

\subsection{A 16.3 day periodic activity window}
\label{sec:16p3days}

We find that FRB 180916.J0158$+$65 is best explained in our model if the QN occurred in the disk of its host galaxy (i.e. high $n_{\rm amb.}^{\rm ns}$) and its chunks bursting in its outskirts (the halo or the IGpM). I.e. FRB 180916.J0158$+$65 is a Disk-Born-Halo-Bursting chunks 
 which yields radial repetition with a few days activity window due to the chunks mass distribution in our model. In this picture, a primary chunk is travelling away from the host's disk
  in the low-density halo or IGpM towards the observer. This is a plausible scenario because 
 the host of FRB 180916.J0158$+$65 is a spiral seen nearly face-on (\citealt{marcote2020}) so we could be seeing the primary
 heading towards us away from its birth place in the disk. We recall that the CSE frequency and all other quantities 
 (in particular its 16.3 day period) which depend
 on the chunk's plasma frequency were set by where the chunk was born; here the disk of the spiral galaxy.

\begin{figure*} 
\centering
\includegraphics[scale=0.7]{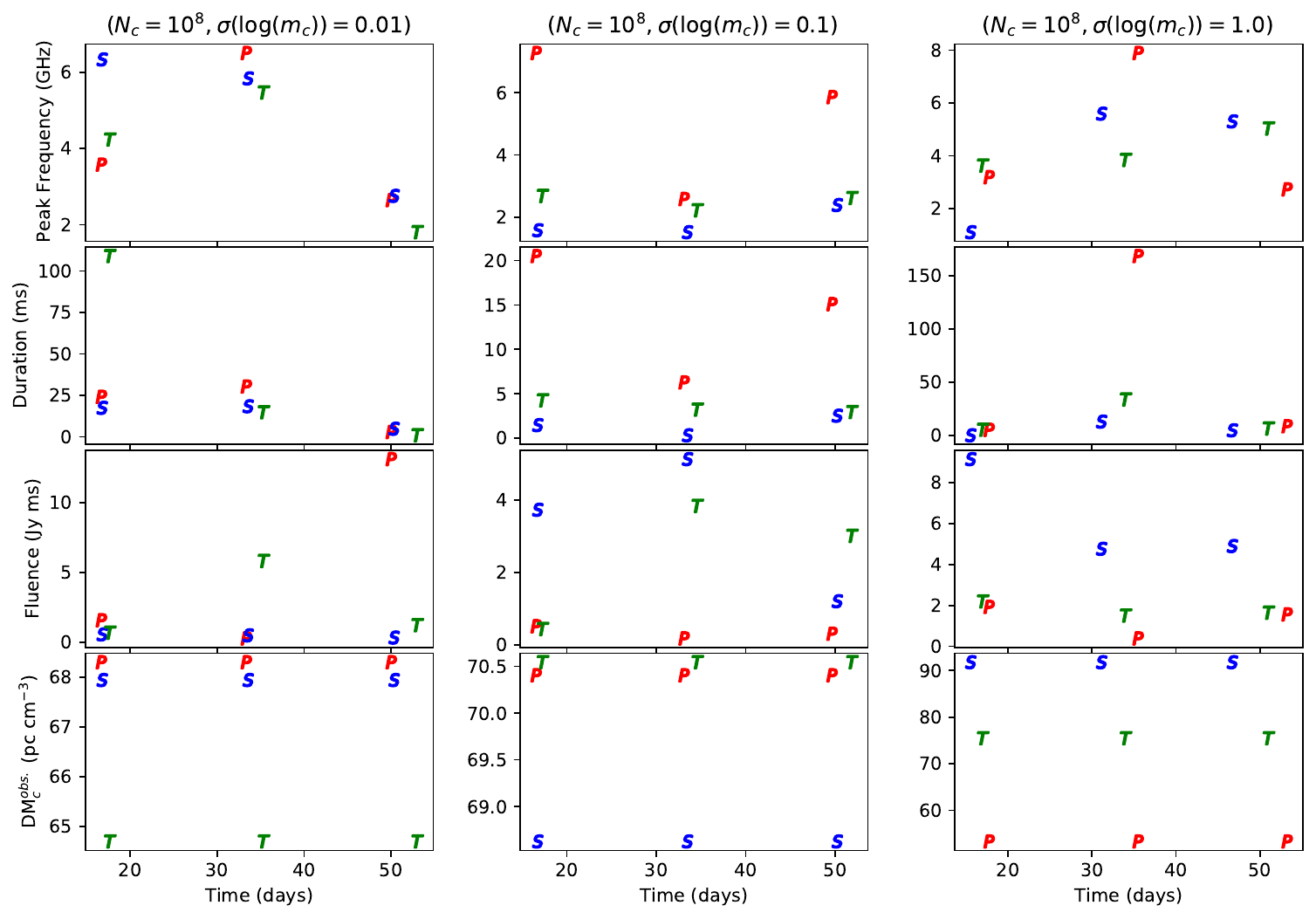}
\caption{An $\sim 16.3$ days repeater in our model with parameters
given in Table \ref{Table:16p3FRB}. The three columns are for $\sigma (\log{(m_{\rm c})})=0.01, 0.1$ and 1.0 from left to right.  Other parameters were set to fiducial values with their log-normal distributions when applicable (see \S \ref{sec:multiple-chunks}). Three BI-WI episodes are shown.}
\label{fig:Periodicity-fiducial-values-10^10-16.3days}
\end{figure*}

FRB 180916.J0158$+$65 (at redshift $z=0.0337$) with its 16.3 day periodic activity window can be reproduced using the parameters listed in Table \ref{Table:16p3FRB}.  Other parameters were kept to their fiducial mean values and varied following a log-normal
distribution as before.  The three columns in Figure \ref{fig:Periodicity-fiducial-values-10^10-16.3days} are for $\sigma (\log{(m_{\rm c})})=0.01, 0.1$ and 1.0 from left to right, respectively.  Here, the mass of the P, S and T chunks, changes from one column to another due changes
 in  $\sigma (\log{(m_{\rm c})})$ which induces  changes in chunks' ${\rm DM}_{\rm c}^{\rm obs.}\propto m_{\rm c}^{2/5}$ from one column to another. A narrow chunk mass distribution is necessary to reproduce properties of the observed 16.3 day FRB with  
 $N_{\rm c}=10^8$ so that $f(\Gamma_{\rm c},\theta_{\rm c})\sim 1$ up to at least the tertiary chunks. 
  Because $\Delta t_{\rm BI-WI}^{\rm obs.}\propto \nu_{\rm CSE, p}^{\rm obs.\ -2}$ (see Eq. (\ref{eq:key-equations})), 
 the radial repetition is chromatic with the higher frequency FRBs arriving first. 
  The few days activity window is due to the chunks mass distribution and becomes narrower as $\sigma (\log{(m_{\rm c})})$ decreases.
   In comparison, the angular time delay between chunks is  $\Delta t_{\rm ang.}^{\rm obs.}\sim 8.8$ s. 
 
 For $\sigma (\log{(m_{\rm c})})\le 0.1$, we have DM$_{\rm c}^{\rm obs.}\simeq 70$ pc cm$^{-3}$ (the same for all chunks) while DM$_{\rm IGM}=z\times 900\simeq 30$ pc cm$^{-3}$. By taking DM$_{\rm MW} = 150$ pc cm$^{-3}$ and DM$_{\rm host}= 100$ pc cm$^{-3}$ we get a  total DM of DM$_{\rm T}\sim 350$ pc cm$^{-3}$ for all chunks. This number agrees with the measured value of 349.5 pc cm$^{-3}$ in 
 FRB 180916.J0158$+$65   (\citealt{chime2019}).  
 
 The maximum distance travelled per radial repeat by a QN chunk in the halo of its host galaxy is $\Delta r_{\rm BI-WI}^{\rm ns}\sim 1.8$ kpc. According to our model, FRB 180916.J0158$+$65 will be active for $\frac{R_{\rm halo}}{\Delta r_{\rm BI-WI}^{\rm ns}}\times 16.3\ {\rm days}$ 
 which is just over a year for $\sim 50$ kpc halo.  However, a much longer activity is possible if the chunk is travelling in the IGpM. 
 Furthermore, the $\Delta r_{\rm BI-WI}^{\rm ns}\sim 1.8$ kpc is an overestimate and can be much smaller (e.g.
 for $\alpha_{\rm diff.}<<1$; see Appendix \ref{appendix:radial-periodicity}) in
 which case the activity can last for many years.  Nevertheless, we acknowledge that explaining variations smaller than 0.1 pc cm$^{-3}$ with our current model poses a challenge.  By the time the chunk crosses the entire halo of its galaxy, the change in DM due to propagation
 should not exceed $\sim 10$ pc cm$^{-3}$ for a $R_{\rm halo}\sim 10$ kpc
 and $n_{\rm halo} \sim 10^{-3}$ cm$^{-3}$. It may even be less if the halo's density is lower or if the chunks are travelling in the IGpM along the observer's line-of-sight.

\begin{figure} 
\centering
\includegraphics[scale=0.3]{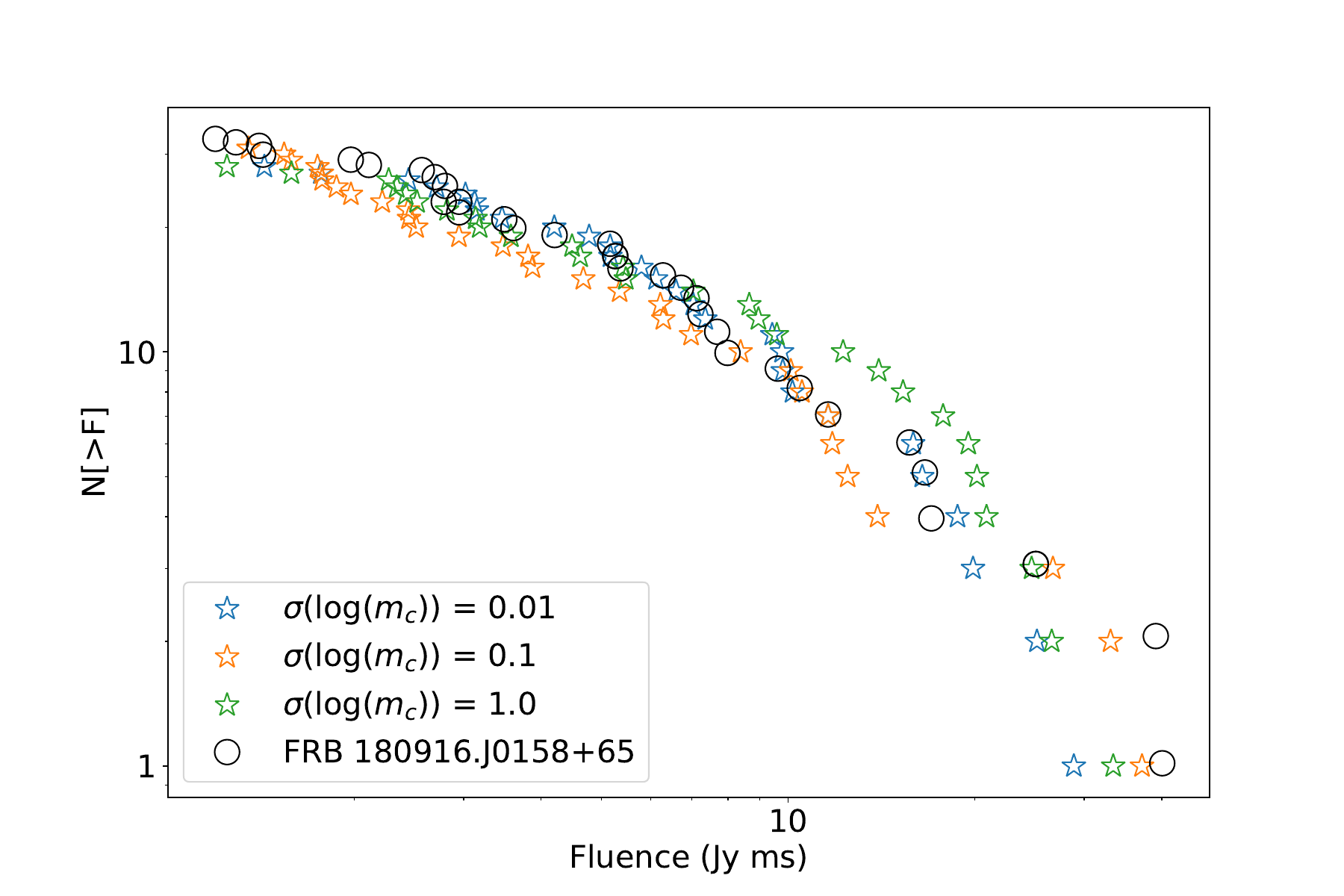}
\caption{The cumulative count distribution of burst fluences derived from the 16.3 day-FRB simulation
in our model (stars) compared to that of CHIME's FRB 180916.J0158$+$65 data (open circles).}
\label{fig:cumulativeSimBaseband16p3}
\end{figure}

Figure \ref{fig:cumulativeSimBaseband16p3} shows the cumulative count distribution of burst fluences derived from our simulations 
(taking into account one primary, six secondaries and twelve tertiaries)
compared to that of FRB 180916.J0158$+$65 Baseband data (\citet{marcote2020}). Our model shows a slightly better agreement
with the $\sigma (\log{(m_{\rm c})})\le 0.01$ case. However when considering more BI-WI episodes, the stochasticity in plasma parameters
washes out the $\sigma (\log{(m_{\rm c})})$ effect and on average good fits can be obtained for all values of $\sigma (\log{(m_{\rm c})})$. Allowing for $\alpha_{\rm diff.}$ to vary between BI-WI episodes increases the width of the periodic activity window 
 but does not change our overall findings as described in Figures \ref{fig:Periodicity-fiducial-values-10^6},
 \ref{fig:Periodicity-fiducial-values-10^10} and  \ref{fig:Periodicity-fiducial-values-10^10-16.3days}

\section{Discussion and testable predictions}
\label{sec:discussion}

We start by discussing some distinctive features of our model:

\begin{itemize}[noitemsep]

\item \label{itemSpatialDistribution}{\bf The FRB spatial distribution}:  We first recall the three key distances in our model (see Figure \ref{fig:single-chunk}): 

\begin{itemize}
\item  $d_{\rm QN}^{\rm ns}\simeq 3.1\ {\rm kpc}\ v_{\rm kick, 300} \tau_{\rm QN, 7}$ which is the mean distance travelled by the NS before it experiences the QN phase. 
Here, $v_{\rm kick}$ and $\tau_{\rm QN}$ are in units of 300 km s$^{-1}$ and $10^7$ years, respectively.

Taking an average viewing angle with respect to a center of a galaxy of $\cos{(\theta_{\rm galaxy})}\sim 2/3$, the typical projected offset  for a QN is  $2\ {\rm kpc}$. 

\item $d_{\rm c}^{\rm ns}\simeq 28.7\ {\rm pc}\times \Gamma_{\rm c, 2.5}^{-1/5}m_{\rm c, 22.3}^{1/5}n_{\rm amb., -1}^{\rm ns\ -3/5}$
given by Eq. (\ref{eq:dc}) which is  the distance travelled by a chunk  until it becomes  collisionless. 
In our model, the NS travels in any direction in the galaxy while the FRB (from the QN chunks) travels directly to the observer because of the $1/\Gamma_{\rm c}$ beaming.  
 It implies an  $1/\Gamma_{\rm c}=10^{-2.5}$ radians 
 angular offset from the line-of-sight.  However, even for the low galactic halo density the resulting
 angular offset can be neglected with $d_{\rm c}^{\rm ns}/\Gamma_{\rm c} << d_{\rm QN}^{\rm ns}$. 

\item $\Delta r_{\rm BI-WI}^{\rm ns}\sim 28.7\ {\rm kpc}\times \frac{1}{\Gamma_{\rm c, 2.5}^{7/5}m_{\rm c, 22.3}^{1/10}n_{\rm amb., -1}^{\rm ns\ 6/5}}$ the maximum distance travelled between successive BI-WI episodes given by Eq. (\ref{eq:distance}), defining radial repetition in our model.

 Chunks from QNe occurring in the halo (i.e. $n_{\rm amb.}^{\rm ns}\sim 10^{-3}$ cm$^{-3}$) will leave their galaxy before they repeat because $\Delta r_{\rm BI-WI}^{\rm ns}\sim 7.2$ Mpc.  Within the galactic disk, chunks from QNe occurring in the WIM (i.e. $n_{\rm amb.}^{\rm ns}\sim 0.1$ cm$^{-3}$) will repeat only once they are in the halo after travelling a distance of $\sim 28.7$ kpc from the QN site. On the other hand, chunks from QNe occurring in the dense neutral medium with $\Delta r_{\rm BI-WI}^{\rm ns}< 1$ kpc will likely repeat while still travelling in the disk and may show ``radial periodicity" once they enter the homogeneous galactic halo where $L_{\rm halo}>> \Delta r_{\rm BI-WI}^{\rm ns}$. We argued
 in \S \ref{sec:16p3days} that FRB 180916.J0158$+$65 is one such candidate where the QN occurred in the galactic disk and we are detecting CSE (i.e. FRBs) from a few of its chunks travelling towards us in the halo of its galaxy or in the IGpM.
\end{itemize}

Events such as  FRB 20240209A which is $\sim 40$ kpc from the center of its host galaxy (\citealt{shah2024}) can be accounted for in our model by QNe from NSs with high kick velocity and/or with longer $\tau_{\rm QN}$ delay. 
For extreme cases, the QN will occur outside the galaxy in which case the associated FRBs will be one-offs and  will appear host-less and progenitor-less.  A more rigorous analysis would require using Monte Carlo simulations by randomly varying $\tau_{QN}$ and $v_{\rm kick}$ 
while considering the two (birth and bursting) environments 
and compare the resulting spatial distribution of QNe in a galaxy with radial offsets of FRBs showing galaxy association (\citealt{aggarwal2021,shannon2024}).

\item \label{itemRate}{\bf The overall FRB rate}: In a galaxy like ours, roughy $1$\% of all neutron stars formed  over a Hubble time experience a QN episode (\citealt{ouyed2020}).
With a core-collapse supernova (ccSN) rate of  $10^{-2}$ per year per high mass ($L_{\star}$) galaxy, about $\sim 10^{-2}\ {\rm yr}^{-1}\times 10^{10}\ {\rm yr}= 10^8$ NSs have formed with $\sim 10^6$ NSs as QN candidates. 
 The QN rate in our model per year per galaxy is thus 
$10^6/10^{10}\ {\rm yr}= 10^{-4}\ {\rm yr}^{-1}$ per galaxy or $\sim 2.7\times 10^{-7}\ {\rm day}^{-1}$ per galaxy. Over the whole sky (with $\sim 10^{11}$ galaxies) we get a rate of $\sim 2.7\times 10^{4}$ per day. The intrinsic FRB rate is much higher by $N_{\rm c}$; but of these only a fraction are observed in any given direction because of beaming. For $\Gamma_c=10^{2.5}$, the beaming factor is $\sim  10^{-5}$ but also depends on how sensitive the detector is. I.e. fewer bursts seen at larger distance from the observer. We used $10^{11}$  galaxies to estimate the QN rate but what matters is the number of galaxies within the upper detection distance of FRBs. Nevertheless, within an order of magnitude we arrive at an FRB rate  which is similar to the observed rate.  A similar rate was derived in  \citet{ouyed2021}
 based on a more rigorous calculation. 
 
\item \label{itemRateRepeaters}{\bf The repeating FRBs rate}:  The low $v_{\rm kick}$ NSs would experience the
QN in the densest environment in their host galaxies (i.e. while still in the disk) and should yield radially repeating FRBs. 
These, and taking $v_{\rm kick}$ of a few km s$^{-1}$,  would
correspond to a small percentage of the total population of FRBs. 
 
\item \label{itemFRBsinBinaries}{\bf The binary connection}: Quark deconfinement in the core of a NS (and the ensuing QN) could also occur via mass-accretion onto the NS from a companion (e.g. in Low-MassX-ray Binaries; see \citet[and references therein]{ouyed2017b}). Some FRBs should be associated  with environments harbouring  massively accreting NSs; e,g, in GCs as in FRB 20200120E (\citealt{bhardwaj2021,kirsten2022}). Even though the FRB would occur well outside the globular cluster, it would be projected in front of it because its motion is along the line-of-sight to the globular cluster.\\
  
If FRB 20200120E is a QN chunk exiting the GC and moving away radially towards us, 
we can estimate its expected apparent angular separation from its birth site. We first note that the chunk will be visible up to $\theta_{\rm c}=1/\Gamma_{\rm c}=10^{-2.5}$ radians from the line-of-sight. Secondly, the time delay between the QN and the FRB is approximately 90 years in the NS frame (see Appendix \ref{appendix:angular-periodicity}). The corresponding radial distance traveled is about $30$ pc, which when projected onto the sky plane gives approximately $30\ {\rm pc}\times \sin{(\theta_{\rm c})}\approx 0.1$ pc. For a GC size of around 1 pc, the FRB would appear near the cluster's center. Thirdly, the angle from the center for the observer depends on the GC distance. For example, a GC at 100 Mpc gives the angle to be $0.1/10^8=10^{-9}$ radians, which is much smaller than an arcsecond ($\simeq 4.8\times 10^{-6}$ radians) and would not be observable. Apparent superluminal motion would not be detectable because the angle is too small to be observed, even with a 90-year baseline for proper motion.

\item \label{itemtheTWOFRBs}{\bf FRB 20180916B and FRB 20121102A}:  FRBs in our model can occur in all types of galaxies and all have in common the CSE emitting collisionless QN chunks and should thus share many properties despite the different hosts. 

FRB 20121102A is best explained in our model as an angular repeater occurring
in a low ambient density medium ($n_{\rm amb.}^{\rm ns}< n_{\rm amb., c}^{\rm ns}$)  with an activity 
lasting for decades as demonstrated in our first paper (see Tables S5, S10 and S11 in \citealt{ouyed2021}).
For FRB 20190608B we favor a Disk-Born-Halo(or IGpM)-Bursting chunk scenario. 
Nevertheless, they should both show similarities in their FRB properties despite the 
 striking difference between the  FRB 20121102A host galaxy (a low-metallicity
dwarf at $z = 0.193$; \citealt{chatterjee2017}) and that of  FRB 20190608B (a spiral at
$z = 0.0337$; \citealt{marcote2020}).

\item \label{itemAngularVSRadial}{\bf Angular vs radial repeaters}: (i) Angular repeaters involve chunks from the same QN but at different viewing angles. For angular repetition to occur, the QN needs to be closer in redshift. This proximity is necessary for the fluence (with $\mathcal{F}\propto f(\Gamma_{\rm c},\theta_{\rm c})^{-2}$) from peripheral chunks to be within detectors' sensitivity.  Angular repeaters are characterized by a lower DM$_{\rm T}$ due to a reduced contribution of intergalactic medium (IGM) DM and a lower DM$_{\rm c}^{\rm obs.}\propto f(\Gamma_{\rm c},\theta_{\rm c})^{-1}$. There is no change in the total DM in this case; 
 (ii) Radial repeaters occur when single chunks experience multiple CSE episodes as they travel radially away from the QN in the ambient plasma. Because  $n_{\rm amb.}^{\rm ns}>n_{\rm amb., c}^{\rm ns}$, the DM 
 of a radially repeating  chunk on average will exceed that of an angular repeater. Also, because $\theta_{\rm c}\sim 0$, a radial repeater can
 be detected at larger redshifts.  We add that in order to experience many CSE episodes, repeaters likely occur in the halo of their hosts
 or outside where the ambient density is extremely small; Disk-Born-Halo(or IGpM)-Bursting chunks. This should be associated with DM$_{\rm host}\sim 0$ and a very small change in DM.

 \item \label{itemSpectrum}{\bf The  spectrum}: In our model, it is a convolution of the CSE spectrum per bunch (a boosted synchrotron spectrum in the  frequency range $\nu_{\rm p, e} \le \nu \le \nu_{\rm CSE, p}$) and the bunch size distribution $\lambda_{\rm b}$
 with $\nu_{\rm CSE, p}=c/\lambda_{\rm b}$.  For a given chunk, and 
  in the simplest possible picture, all bunches have the same size and grow as $\lambda_{\rm b}(t)\propto (1+t/t_{\rm m-WI})$. 
  A preliminary analysis is given in Appendix \ref{appendix:spectrum} with an example spectrum
   given in Figure \ref{fig:CSE-spectrum}. However, the final spectrum is more complex when considering a distribution of bunches
    at any given time and their evolution due to turbulent filament merging. Furthermore, 
  while the number of electrons per bunch is constant for a given chunk (see Eq. (\ref{eq:Ne}) in Appendix \ref{appendix:spectrum}),
  the distribution of the electron Lorentz factor $\gamma_{\rm e}$ will vary from bunch-to-bunch within a chunk.
    The underlying  CSE spectra is more likely to be smeared out and would lose its underlying
    synchrotron shape (i.e. $\nu^{1/3}$ for $\nu < \nu_{\rm CSE, p}$) and should instead
    carry the signature of the turbulent merging of the WI filaments.
   A detailed analysis is beyond the scope of this paper  and will be done elsewhere.
   
   \item \label{itemDMpropagation}{\bf The  propagation DM}: Measured change in DM for repeaters is typically $\pm 3$ pc cm$^{-3}$ (e.g. \citealt{hu2023}).
   In our model, repeaters should have slowly declining DM due to propagation (\S \ref{sec:propagation-DM}). The observed DM limits imply a low $n_{\rm ionized}^{\rm ns}$ (hinting at galactic halos, IGpM or IGM) or short $\Delta r_{\rm BI-WI}^{\rm ns}$ which could be caused by high $n_{\rm amb.}^{\rm ns}$ (i.e. QNe occur mainly in galactic disks) and/or faster magnetic field decay after WI saturation (see Appendix \ref{appendix:radial-periodicity}).

 \end{itemize}
 
 We end by listing a few testable predictions:
 
 \begin{itemize}[noitemsep]
 
 \item \label{itemProbes}{\bf FRBs as probes of ionized media in the universe}: Because the CSE luminosity scales as $n_{\rm ionized}$ (see Eq. (\ref{eq:chunk-frame}) in Appendix \ref{appendix:CSE}), as a radially repeating chunk encounters density jumps it should be reflected in the fluence. When the jumps are small such variations may be 
masked by the stochasticity related to the plasma parameters $\beta_{\rm WI}$ and  $\zeta_{\rm BI}$. 
However, there are instances where the jump in $n_{\rm amb.}^{\rm ns}$ may be important enough that it can reflect directly
on the fluence: (i) WIM ($n_{\rm amb.}^{\rm ns}\sim 10^{-1}$ cm$^{-3}$) to Galactic-halo ($n_{\rm amb.}^{\rm ns}\sim 10^{-3}$ cm$^{-3}$); (ii) Galactic-halo ($n_{\rm amb.}^{\rm ns}\sim 10^{-3}$ cm$^{-3}$) to inter-group  ($n_{\rm amb.}^{\rm ns}\sim 10^{-5}$ cm$^{-3}$);
(iii) inter-group  ($n_{\rm amb.}^{\rm ns}\sim 10^{-5}$ cm$^{-3}$) to IGM ($n_{\rm amb.}^{\rm ns}\sim 10^{-7}$ cm$^{-3}$);
(iv) The most important jump in fluence would be when a CSE emitting chunk in the halo exits the galaxy and enters the IGM. While the
CSE frequency should remain the same, we expect a dimming in fluence by a factor of $n_{\rm IGM}/n_{\rm halo}\sim 10^{-4}$.

  We expect FRB 180916.J0158$+$65 to eventually dim as the emitting QN chunk enters  the IGpM (if it is currently emitting in its host's halo) or
 the IGM if it is currently emitting in the surrounding IGpM. In the process, its fluence will decrease by 
 a factor of tens to hundreds while other quantities remain the same. 
 
 \item \label{itemFRBsinDisk}{\bf FRBs in the galactic disk:} In the galactic disk, FRBs occur whenever a collisionless QN chunk enters an ionized medium and may repeat if $L_{\rm ionized}> \Delta r_{\rm BI-WI}^{\rm ns}$. The most likely environment is the  warm ionized medium (WIM;  \citealt{hoyle1963}) which accounts for about half of the disk's volume
 (e.g. \citealt{reynolds1991,ferreire2001,haffner2009}).  These FRBs should in principle be associated with \lbrack CII\rbrack\ and \lbrack NII\rbrack\  in absorption towards bright continuum sources associated with the WIM (e.g. \citealt{persson2014,gerin2015}).
 
\item \label{itemLFRBs}{\bf Long duration FRBs}: The FRB duration in our model is related to that of the WI
filament merging phase with $\Delta t_{\rm CSE}^{\rm obs.}\propto (1/\beta_{\rm WI}^{1/2}-1)\times \zeta_{\rm m-WI}/\nu_{\rm p, e}$.
The distribution in the astrophysical and plasma parameters yields FRBs lasting for minutes as can be seen
from Figures \ref{fig:compare-to-HERTA}, \ref{fig:Periodicity-fiducial-values-10^6} and \ref{fig:Periodicity-fiducial-values-10^10}.

\item \label{itemHostlessSourceless}{\bf ``Hostless and Sourceless/progenitor-less" FRBs}:  
 NSs with extreme kick velocity (e.g. $v_{\rm kick}\sim 10^3$ km s$^{-1}$) and with long time delays to the QN conversion phase (e.g. $\tau_{\rm QN}\sim 10^8$ years)  would have left their host galaxy by the time they become FRB sources. Thus we predict,
 rare but possible hostless and progenitor-less one-off FRBs in the intra-cluster or inter-galactic medium.
  
  These may be confirmed by the discovery of bright (i.e. nearby) FRBs with no identifiable host nor progenitor despite deep imaging. 
 A hostless FRB in our model would have DM$_{\rm host}=0$ but it may have an important intrinsic chunk DM,
DM$_{\rm c}^{\rm obs.}$. I.e. hostlessness is not synonymous with high-redshift (i.e. due detection limit). 

\item \label{itemChunkDM}{\bf Chunk intrinsic DM in FRB data}: From Eq. (\ref{eq:key-equations-2}) we observe that DM$_{\rm c}^{\rm obs.}\propto m_{\rm c}^{2/5}n_{\rm amb.}^{\rm ns\ 4/5}\Gamma_{\rm c}^{13/5}$. 
I.e. when massive chunks are born and transition into a collisionless state within a dense environment, (e.g. $n_{\rm amb.}>n_{\rm amb., c}$), the chunks DM contribution amounts to 
hundreds of pc cm$^{-3}$,  if not more, particularly for higher values of $\Gamma_{\rm c}$. We predict the detection of a precisely localized and progenitorless FRB with DM far exceeding DM$_{\rm IGM}+{\rm DM}_{\rm MW}$.  This also applies for nearby FRBs 
 where the host's DM is well constrained by $H_{\alpha}$ and which show a cosmic DM far above the 
one expected using the Macquart relation and its scatter.

\item \label{itemzDistribution}{\bf The redshift distribution}:  The typical time delay between the SN and the QN (i.e. $\tau_{\rm QN}\sim$ tens of  millions of years) implies that the  redshift distribution of FRBs should closely follow that of the star formation rate, consistent with recent modelling efforts (e.g., \citealt{james2022,shin2023}).

\end{itemize}

\section{conclusion}
\label{sec:conclusion}

The agreement between our model and FRB data suggests that QN events may be behind this phenomenon. I.e. 
FRBs are CSE from relativistic collisionless QN chunks interacting with various ionized media in the universe.  In our model, FRBs occur millions of years after the birth of the NS experiencing the QN event. Furthermore, the QN ejecta needs to travel a considerable distance from the QN before becoming collisionless. As a result, most FRBs  (and particularly one-offs) are expected to occur in the outskirts (typically offset by a few kpc from the center) of galaxies. NSs undergoing the QN phase outside their host galaxy (e.g. NSs with high $v_{\rm kick}$ and/or long $\tau_{\rm QN}$) will yield FRBs with no identifiable sources or progenitors. FRBs may eventually be discovered in extraordinary locations far from star-forming regions, such as the halos of galaxies, globular clusters, the intra-cluster medium (ICM), and the intergalactic medium (IGM). Some FRBs should be associated with a DM far exceeding expected values because of the intrinsic chunk's DM.

\section*{Acknowledgements}

We thank the referee for insightful comments and suggestions which improved the quality and rigour of our paper.

\section*{Data availability}

The data underlying this article are available in the article.



\begin{appendix}

\section{Ejecta geometry, properties and angular periodicity}
\label{appendix:angular-periodicity}

Figure \ref{fig:multiple-chunks} illustrates the geometry and distribution of chunks as seen by an observer. We assume
a uniformly spaced chunks case here (see Appendix SA in \citealt{ouyed2021} for details and other geometries). For the simplest isotropic distribution of chunks around the center of the QN we have $N_{\rm c}\pi \Delta \theta_{\rm c}^2=4\pi$
with $\Delta \theta_{\rm c}\sim 2/N_{\rm c}^{1/2}$  the angular separation between adjacent chunks and $N_{\rm c}$ the total number of chunks. The number of chunks per ring of index $i_{\rm r}$ is $N(i_{\rm r})= 2\pi (i_{\rm r}\Delta \theta_{\rm c})/\Delta \theta_{\rm c}= 2\pi \times i_{\rm r}$ with $i_{\rm r}=0$ corresponding to the primary chunk so that $N(0)=1$. The total number of chunks up to ring $i_{\rm r, max.}$ is
 $N(i_{\rm r, max.})= \sum_{i=0}^{i_{\rm r, max.}} N(i_{\rm r})= 1+2\pi \sum_{i=1}^{i_{\rm r, max.}} i_{\rm r}= 1+2\pi \times \frac{i_{\rm r, max.}(i_{\rm r, max.}+1)}{2} = 1+\pi \times i_{\rm r, max.}(i_{\rm r, max.}+1)$. Thus, there are roughly 1 primary, $2\pi\sim 6$ secondary and $4\pi\sim 12$ tertiary chunks adding up to a total of $(1+ 6\pi)\sim 19$ chunks. 

 In this geometry the range and mean angle of the primary (P), secondary (S) and tertiary
(T) chunks are $\bar{\theta}_{\rm P}\simeq 4/3N_{\rm c}^{1/2},  \bar{\theta}_{\rm S}\simeq 28/9N_{\rm c}^{1/2}\simeq 2.4\bar{\theta}_{\rm P},
\bar{\theta}_{\rm T}\simeq 49/6N_{\rm c}^{1/2}\simeq 6\bar{\theta}_{\rm P}$. 
The details can be found in Appendix SA in \citet{ouyed2021} where one can show that 
 using the averaged viewing angle for the P, S and T chunks one finds

\begin{align}
f(\Gamma_{\rm c},\bar{\theta}_{\rm P})&\sim 1+ 0.18\frac{\Gamma_{\rm c, 2.5}^2}{N_{\rm c, 6}}\\\nonumber
f(\Gamma_{\rm c},\bar{\theta}_{\rm S})&\sim 1+ 0.97\frac{\Gamma_{\rm c, 2.5}^2}{N_{\rm c, 6}}\\\nonumber
f(\Gamma_{\rm c},\bar{\theta}_{\rm T})&\sim 1+ 6.67\frac{\Gamma_{\rm c, 2.5}^2}{N_{\rm c, 6}}\ .
\end{align}
We see that for $N_{\rm c}> 10^8\Gamma_{\rm c, 2.5}^2$, we have $f(\Gamma_{\rm c},\bar{\theta}_{\rm c})\simeq 1$ up to 
the T chunks.

The average variation in $f(\Gamma_{\rm c},\theta_{\rm c})$ between successive chunks is $\Delta f\sim \frac{1.6}{\pi}\times \frac{\Gamma_{\rm c, 2.5}^2}{N_{\rm c, 6}}$ (Eq. (SA7) in \citealt{ouyed2021}).

The time it takes a QN chunk to become collisionless as it travels in the ambient medium since the QN event is (see Eq. (SB13) and Appendix SB in \citealt{ouyed2021} for details)

\begin{equation}
\label{eq:appendix:tc}
t_{\rm c, ang.}^{\rm obs.}\simeq 6.0\ {\rm hrs}\times (1+z)f(\Gamma_{\rm c},\theta_{\rm c})\times \frac{m_{\rm c, 22.5}^{1/5}}{\Gamma_{\rm c, 2.5}^{11/5}n_{\rm amb, -1}^{3/5}}\ .
\end{equation}
In the NS frame it is $\Gamma_{\rm c}t_{\rm c, ang.}= \Gamma_{\rm c}t_{\rm c, ang.}^{\rm obs.} (1+z)/D(\Gamma_{\rm c},\theta_{\rm c}) \sim 93\ {\rm years}\times m_{\rm c, 22.3}^{1/5}/\Gamma_{\rm c, 2.5}^{1/5}n_{\rm amb.,-1}^{\rm ns\ 3/5}$.

The average angular time separation between chunks, from $ \Delta t_{\rm ang}^{\rm obs.}\propto \Delta f$, is  then 
 \begin{equation}
 \label{eq:Pcang}
 \Delta t_{\rm ang.}^{\rm obs.}\simeq 3.0\ {\rm hrs}\times (1+z)\times \frac{m_{\rm c, 22.5}^{1/5}}{N_{\rm c, 6}\Gamma_{\rm c, 2.5}^{1/5}n_{\rm amb, -1}^{3/5}}\ ,
 \end{equation}
 keeping in mind that  $m_{\rm c}=\frac{{(E_{\rm QN}^{\rm ke})}^{\rm ns}}{N_{\rm c}\Gamma_{\rm c}c^2}\sim 10^{22.3}\ {\rm gm}\times \frac{{(E_{\rm QN, 52}^{\rm ke}})^{\rm ns}}{N_{\rm c, 6}\Gamma_{\rm c, 2.5}}$. Here $(E_{\rm QN}^{\rm ke})^{\rm ns}\simeq 10^{52}$ ergs is the kinetic energy of the QN ejecta which is a small percentage of the total (confinement and gravitational) binding energy released when converting all of the neutrons of a NS to quarks (see \S \ref{sec:intro}).
 
 As the chunk propagates away from the QN and expands, it eventually becomes ionized and 
  collisionless when the electron Coulomb collision length inside the chunk is of the order of 
its radius (see Appendix SB in \citealt{ouyed2021}).  
 The distance travelled by a chunk (in the parent NS frame) until it becomes  collisionless is $d_{\rm c}^{\rm ns}= c\Gamma_{\rm c}\times (D(\Gamma_{\rm c},\theta_{\rm c}) t_{\rm c, ang.}^{\rm obs.}/(1+z))$. I.e. 
 \begin{equation}
 \label{eq:dc}
 d_{\rm c}^{\rm ns}\simeq 28.7\ {\rm pc}\times \Gamma_{\rm c, 2.5}^{-1/5}m_{\rm c, 22.3}^{1/5}n_{\rm amb., -1}^{\rm ns\ -3/5}\ .
 \end{equation}
 
 The chunk's number density, $n_{\rm c}$ and  radius ($R_{\rm c}=(m_{\rm c}/\frac{4\pi}{3}n_{\rm c}m_{\rm H})^{1/3}$ with
$m_{\rm H}$ the hydrogen mass) when it becomes collisionless were  derived in Appendix SA in \citet{ouyed2021}:

\begin{align}
\label{eq:appendix:ncandRc}
n_{\rm c} &\simeq 3.7\times 10^3\ {\rm cm}^{-3}\times m_{\rm c, 22.3}^{1/10} \Gamma_{\rm c, 2.5}^{12/5}  {n_{\rm amb, -1}}^{6/5}\\\nonumber
R_{\rm c}&\simeq 9.3\times 10^{13}\ {\rm cm}\times m_{\rm c, 22.3}^{3/10} \Gamma_{\rm c, 2.5}^{-4/5}  {n_{\rm amb, -1}}^{-2/5}\ .
\end{align} 
We take the proton hadronic cross-section to be $10^{-27}$ cm$^2$ and the chunk opacity $\kappa_{\rm c}=0.1$ cm$^2$ gm$^{-1}$ 
when deriving the equations above (see \citealt{ouyed2021}).
 
  Once collisionless, the chunk remains so (and stops expanding)  since the mean-free-path of an external proton with respect to the chunk's
density always exceeds the chunk's size. Once set,  the density $n_{\rm c}$ (and thus the plasma frequency; $\nu_{\rm p, e}\propto n_{\rm c}^{1/2}$) and the size $R_{\rm c}$ remain fixed as the chunk encounters different ionized media inside and outside its host galaxy.

\section{Coherent Synchrotron Emission (CSE) from the collisionless QN chunks}
\label{appendix:CSE}

\renewcommand\thesection{\Alph{section}}
\renewcommand{\thefigure}{\thesection.\arabic{figure}}
\setcounter{figure}{0} 

\subsection{Weibel current filaments}
\label{appendix:WI-filaments}

The interaction between the ambient plasma protons (the beam) and the chunk's electrons (the background plasma) triggers 
the Buneman Instability (BI) with the generation of electrostatic waves inducing  an anisotropy in the chunk's electron population thermal velocity
by increasing the electron velocity component along the direction of motion of the chunk; i.e. $v_{\perp}<v_{\parallel}$. 
This anisotropy in turns triggers the thermal Weibel Instability (WI) which induces cylindrical 
current filaments along the  direction of motion  of the chunk and the generation of the WI magnetic field which saturates when it is 
 close to equipartition value $B_{\rm c, s}^2/8\pi\sim n_{\rm c}m_{\rm p}c^2$.  The filament's initial width (i.e. its transverse size at t=0) is 
  set by the dominant electron-WI mode with wavelength $\lambda_{\rm F}(t=0)=\lambda_{\rm F, min.}= \beta_{\rm WI}^{1/2}\times c/\nu_{\rm p, e}$ where  $\beta_{\rm WI}=v_{\perp}/v_{\parallel}< 1$ the chunk's thermal anisotropy parameter ($v_{\perp}$ and $v_{\parallel}$
  are the electron's speed in the direction perpendicular and parallel to the beam; see \citealt{ouyed2021} and references therein for details).  Here, $2\pi \nu_{\rm p, e}=\sqrt{4\pi e^2 n_{\rm c, e}/m_{\rm e}}\simeq 9\times 10^3\ {\rm Hz}\times n_{\rm c, 0}^{1/2}$ is the
  chunk's plasma frequency and $n_{\rm c}$ its number density with $n_{\rm c, e}=n_{\rm c}$; $m_{\rm e}$ and $e$ are the electron mass
  and charge, respectively.
  
  In the non-linear regime of the WI, the WI filaments merge and grow in time as $\lambda_{\rm F}(t) = \lambda_{\rm F, min.}\times \left(1+\frac{t}{t_{\rm m-WI}} \right)^{\delta_{\rm m-WI}}$ with $\delta_{\rm m-WI}$ the filament merging index and $t_{\rm m-WI}=\sqrt{\frac{m_{\rm p}}{m_{\rm e}}}\times \frac{\zeta_{\rm m-WI}}{\nu_{\rm p, e}}$  the corresponding characteristic merging timescale (see Appendix SC3 and Eq. (SC11) in \citealt{ouyed2021}). The filaments grow in size until they reach a maximum size $\lambda_{\rm F, max.}=c/\nu_{\rm p, e}$ set by the plasma skin depth.
  
  Theoretical studies and Particle-In-Cell  (PIC) simulations of the
thermal WI instability suggests $\delta_{\rm m-WI}\sim 0.8$ and $\zeta_{\rm m-WI}$ in the thousands (\citealt{medvedev2007,nishikawa2021}; see also \citealt{takabe2023} for a recent review). We adopt $\delta_{\rm m-WI}\sim 1.0$ and $\zeta_{\rm m-WI}=10^5$ as the fiducial values. Together, the WI-BI tandem converts a percentage ($\zeta_{\rm BI}$) of the chunk's kinetic energy to magnetic field energy and to accelerating the chunk's electrons to relativistic speeds. These then cool and yield the CSE in our model (see Appendix SC in \citealt{ouyed2021}).

\subsection{Bunch properties}
\label{appendix:Neb-and-NbT}

The electrons that have been accelerated by the BI are often spatially bunched in form of phase space holes (\citealt{roberts1967,dieckmann2007}). 
Electron bunching occurs in the peripheral region around the filaments  with bunch size $\lambda_{\rm b}=\delta_{\rm b}\lambda_{\rm F}$ where $\delta_{\rm b}<1$ (see details in section SD and Figure S3 in  \citealt{ouyed2021}). 
The number of electrons per bunch is $N_{\rm e, b}= (2\pi\lambda_{\rm F}\lambda_{\rm b})\times (2R_{\rm c})\times n_{\rm c}$.  
 Because $\lambda_{\rm F, min.}=\beta_{\rm WI}^{1/2}/\nu_{\rm p, e}$  and $n_{\rm c}\propto \nu_{\rm p, e}^2$, the number 
 of electrons per bunch is (see also  \S SD2 and Eq. (SD11) in \citet{ouyed2021}) 
 \begin{equation}
 \label{eq:Ne}
 N_{\rm e, b}(t)\simeq 8.8\times 10^{29}\times \delta_{\rm b}\beta_{\rm WI}R_{\rm c, 15}\times \left(1+\frac{t}{t_{\rm m-WI}} \right)^{2\delta_{\rm m-WI}}\ .
 \end{equation}

 The maximum total number of filaments ($N_{\rm b, T}=(R_{\rm c}/\lambda_{\rm F})^2$) decreases over time as
 \begin{equation}
  \label{eq:Nb}
 N_{\rm b, T}= 9\times 10^{16} \times \frac{R_{\rm c, 15}^2n_{\rm c,0}}{\beta_{\rm WI}}\times \left(1+\frac{t}{t_{\rm m-WI}} \right)^{-2\delta_{\rm m-WI}}\ .
 \end{equation}
The exact  number of filaments is hard to quantify but it is likely to be much less than the maximum value given above. 

\subsection{CSE frequency and duration}
\label{appendix:CSE-frequency-and-duration}

\begin{figure*}
\centering
\includegraphics[scale=0.5]{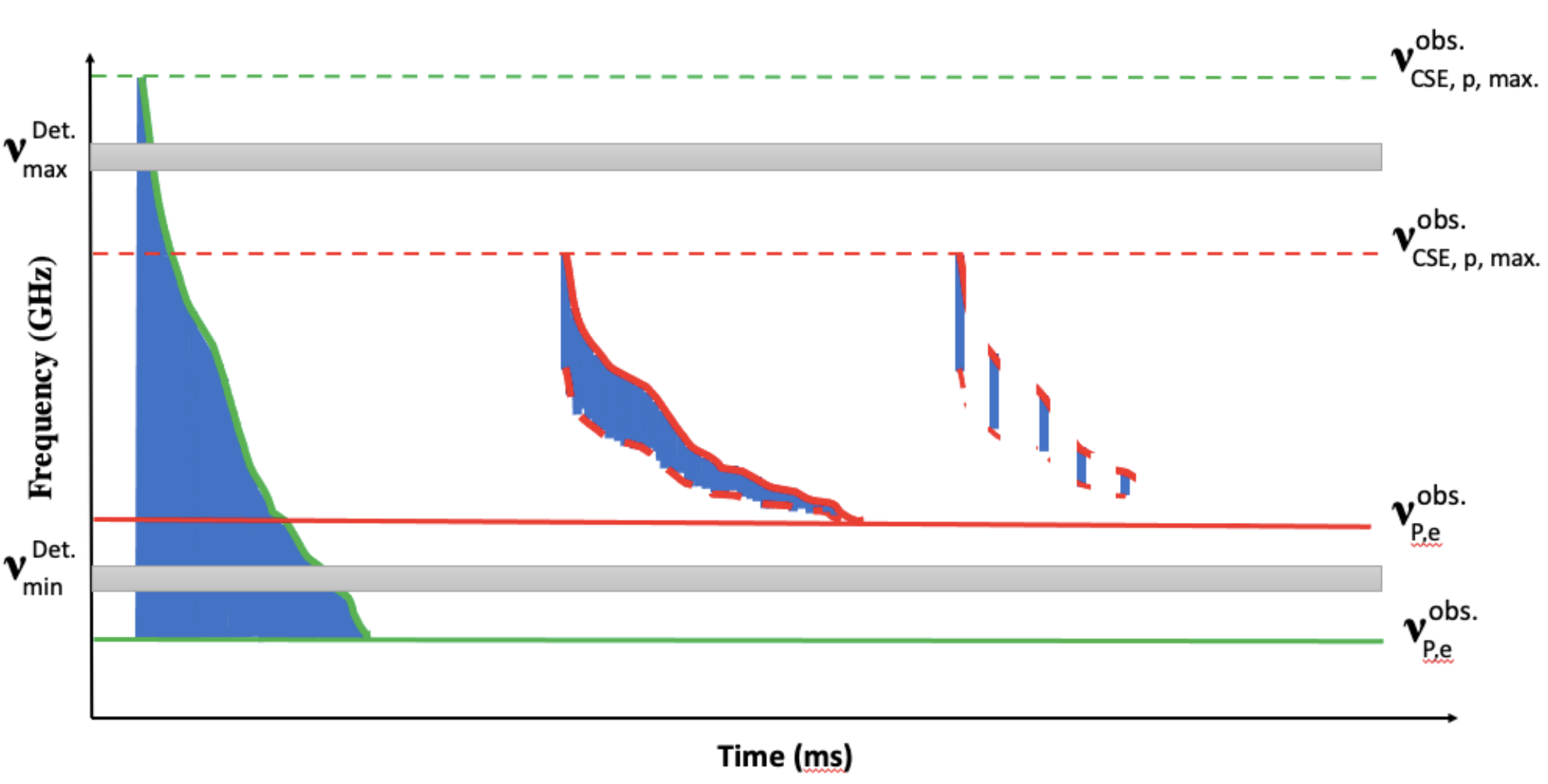}
\includegraphics[scale=0.5]{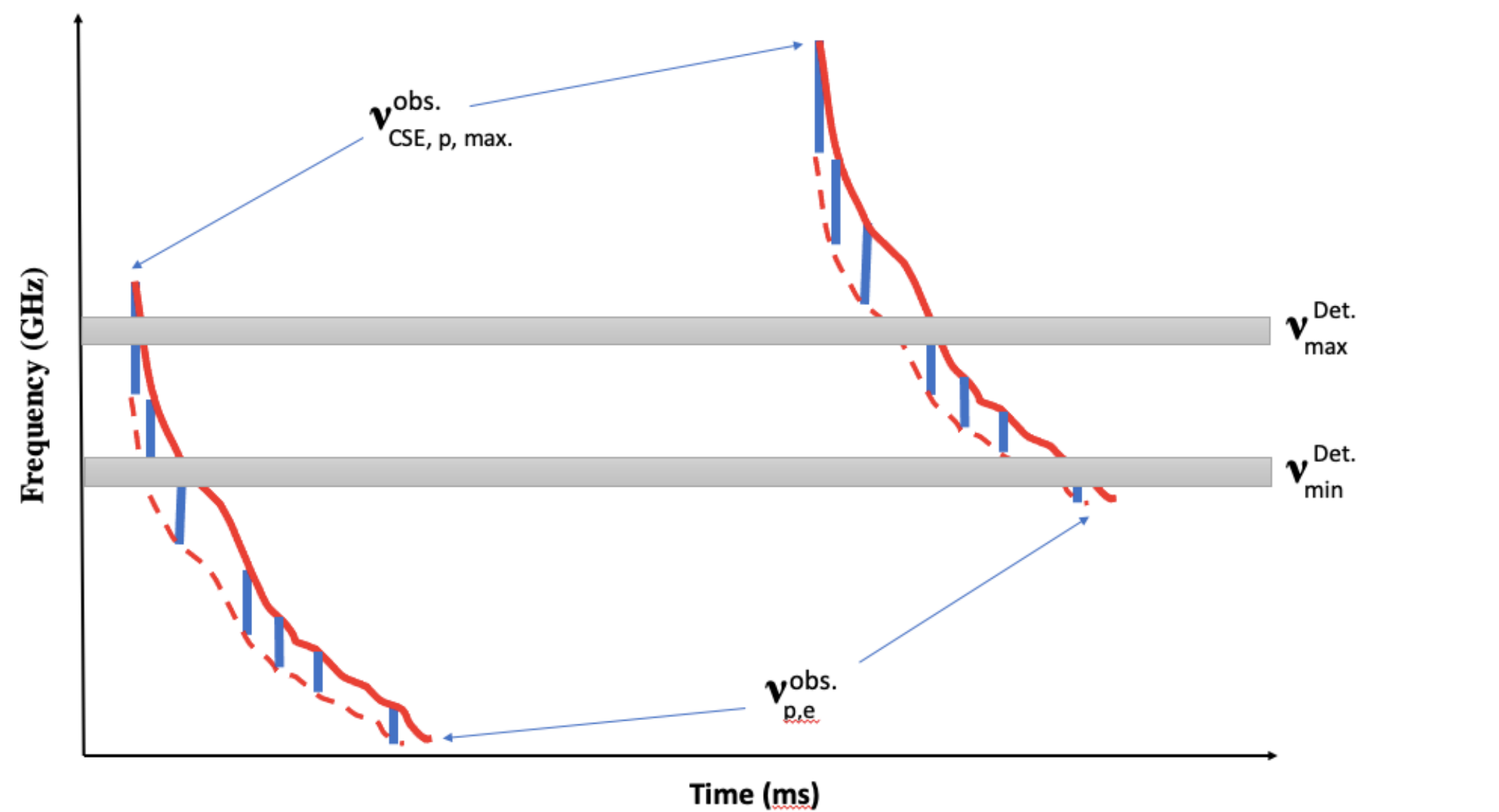}
\caption{{\bf Top panel}: Examples of CSE  peak frequency drift in our model. The solid line depicts the $\nu_{\rm CSE, p}^{\rm obs.}(t)\propto (1+ t^{\rm obs.}/t_{\rm m-WI}^{\rm obs.})^{-\delta_{\rm m-WI}}$
 trend due to WI filament (and thus of electron bunches) merging and growing in size; $\delta_{\rm m-WI}>0$. 
 The dashed red lines illustrate detector's sensitivity cut-off (i.e. with lower frequency emission below detector's sensitivity). The drift ceases when the filament size is of the order of the plasma skin depth (i.e. when $\nu_{\rm CSE, p}^{\rm obs.}=\nu_{\rm p, e}^{\rm obs.}$). The leftmost diagram is for a flat spectrum with drifting across the detector's band of width $(\nu_{\rm max.}^{\rm Det.}-\nu_{\rm min.}^{\rm Det.})$. At any given time during the drifting, emission is in the frequency range $\nu_{\rm p, e}^{\rm obs.} \le \nu_{\rm CSE}^{\rm obs.} \le \nu_{\rm CSE, p}^{\rm obs.}(t^{\rm obs.})$. The middle diagram shows a power-law spectrum with the minimum and maximum CSE frequencies
 within the detector's band. The right diagram shows the gaps in CSE (i.e. intermittent emission) due to the stochastic nature
 of WI filament (and thus bunch) merging and growth which gives rise to a  ``sad trombone effect" as discussed in \S \ref{sec:sadtrombone}. 
 {\bf Bottom panel}: All FRBs are intermittent due to filament merging. The solid and dashed lines are frequency bounds while blue bands are the emission. The left burst illustrates a non-repeating FRB and the right burst illustrates a repeating FRB. Repeaters in our model are CSE from chunks born in a high density ambient environment with $n_{\rm amb.}^{\rm ns}>n_{\rm amb., c}^{\rm ns}$ (see Eq. (\ref{eq:nambc}) and Figure \ref{fig:WIM-and-Halo} for an illustration). Their density ($n_{\rm c}\propto n_{\rm amb.}^{\rm ns\ 6/5}$) and thus their plasma frequency 
 ($\nu_{\rm p, e}\propto n_{\rm c}^{1/2}$) is higher than that of one-offs; here the right burst is the left one shifted upwards. Detectors on earth will catch the broader end of the pulse (i.e. of the filament merging process) which implies longer duration FRBS for repeaters according to our model and a higher probability of capturing a ``sad trombone" effect.}
\label{fig:drifting}
\end{figure*}

The relativistic electrons in the bunch emit coherently in the frequency range $\nu_{\rm p, e}\le  \nu_{\rm CSE}\le \nu_{\rm CSE, p}(t)$ with the peak CSE frequency given as $\nu_{\rm CSE, p}(t)= c/\lambda_{\rm b}(t)$.
 The bunches grow in size with the merging WI current filaments which induces 
 a decrease in the peak CSE frequency (and the narrowing of the emitting band) as

 \begin{equation}
 \label{eq:nu-drift-appendix}
\nu_{\rm CSE, p}(t) = \frac{c}{\delta_{\rm b}\lambda_{\rm F}(t)}= \nu_{\rm CSE, p, max.}\times  \left(1+\frac{t}{t_{\rm m-WI}} \right)^{-\delta_{\rm m-WI}}\ .
\end{equation}
The above profile is illustrated in Figure \ref{fig:drifting} under different scenarios. 
 
  At $t=0$, before filament merging starts, a given bunch emits in the frequency range $\nu_{\rm p, e}\le \nu_{\rm CSE} \le \nu_{\rm CSE, p, max.} =\nu_{\rm p, e}/\delta_{\rm b}\beta_{\rm WI}^{1/2}$. I.e.
  
  \begin{equation}
  \frac{\nu_{\rm CSE, p, max.}}{\nu_{\rm p, e}}= \frac{1}{\delta_{\rm b}\beta_{\rm WI}^{1/2}}\ .
  \end{equation}
  
  The minimum CSE peak frequency is set by the maximum bunch size from merging $\lambda_{\rm b, max.}= \delta_{\rm b}\lambda_{\rm F, max.}$. This gives $\nu_{\rm CSE, p, min.}= c/\lambda_{\rm F, max.}=\nu_{\rm p, e}/\delta_{\rm b}$ so that 

 \begin{equation}
  \frac{\nu_{\rm CSE, p, min.}}{\nu_{\rm p, e}}= \frac{1}{\delta_{\rm b}}\ .
  \end{equation}
 
I.e. while the lower limit (i.e. the chunk's plasma frequency) is constant for a given chunk, the CSE frequency
  band becomes narrower as the peak frequency slides down from $\nu_{\rm p, e}/\delta_{\rm b}\beta_{\rm WI}^{1/2}$
  to $\nu_{\rm p, e}/\delta_{\rm b}$. The CSE duration is found by setting $\nu_{\rm CSE, p}(\Delta t_{\rm CSE})=\nu_{\rm CSE, p, min.}$,

\begin{align}
\label{eq:dt-chunk-frame}
\Delta t_{\rm CSE}&= t_{\rm m-WI}\times \left( \left(\frac{\nu_{\rm CSE, p, max.}}{\nu_{\rm CSE, p, min.}}\right)^{1/\delta_{\rm m-WI}}-1\right)\ .
\end{align}
which for $\delta_{\rm m-WI}=1$ means 
\begin{equation}
\label{eq:dt-CSE}
\Delta t_{\rm CSE} = t_{\rm m-WI}\times \left(\frac{1}{\beta_{\rm WI}^{1/2}}-1\right) \ .
\end{equation}

Hereafter, we take $\delta_{\rm b}=1$ (i.e. $\lambda_{\rm b}=\lambda_{\rm F}$) which means that we are equating the bunch size to that of the Weibel filament's radius. I.e. the CSE peak frequency drifts from $\nu_{\rm CSE, p, max.}=\nu_{\rm p, e}/\beta_{\rm WI}^{1/2}$ to  ${\nu_{\rm CSE, p, min.}}=\nu_{\rm p, e}$. 
 This simplification affects the width of the emission band while still capturing most model properties, including the sad-trombone effect (see \S \ref{sec:sadtrombone}).
  Key quantities are not impacted since $\delta_{\rm b}$ cancels out when taking ratios like $\nu_{\rm CSE, p, max.}/\nu_{\rm CSE, p, min} = (1/\beta_{\rm WI}^{1/2} - 1)$ and $\Delta \nu_{\rm CSE, p}/\nu_{\rm CSE, p, min} = (\nu_{\rm CSE, p, max.} - \nu_{\rm CSE, p, min})/\nu_{\rm CSE, p, min} = 1/\beta_{\rm WI}^{1/2}$.

\begin{figure*}
\centering
\includegraphics[scale=0.6]{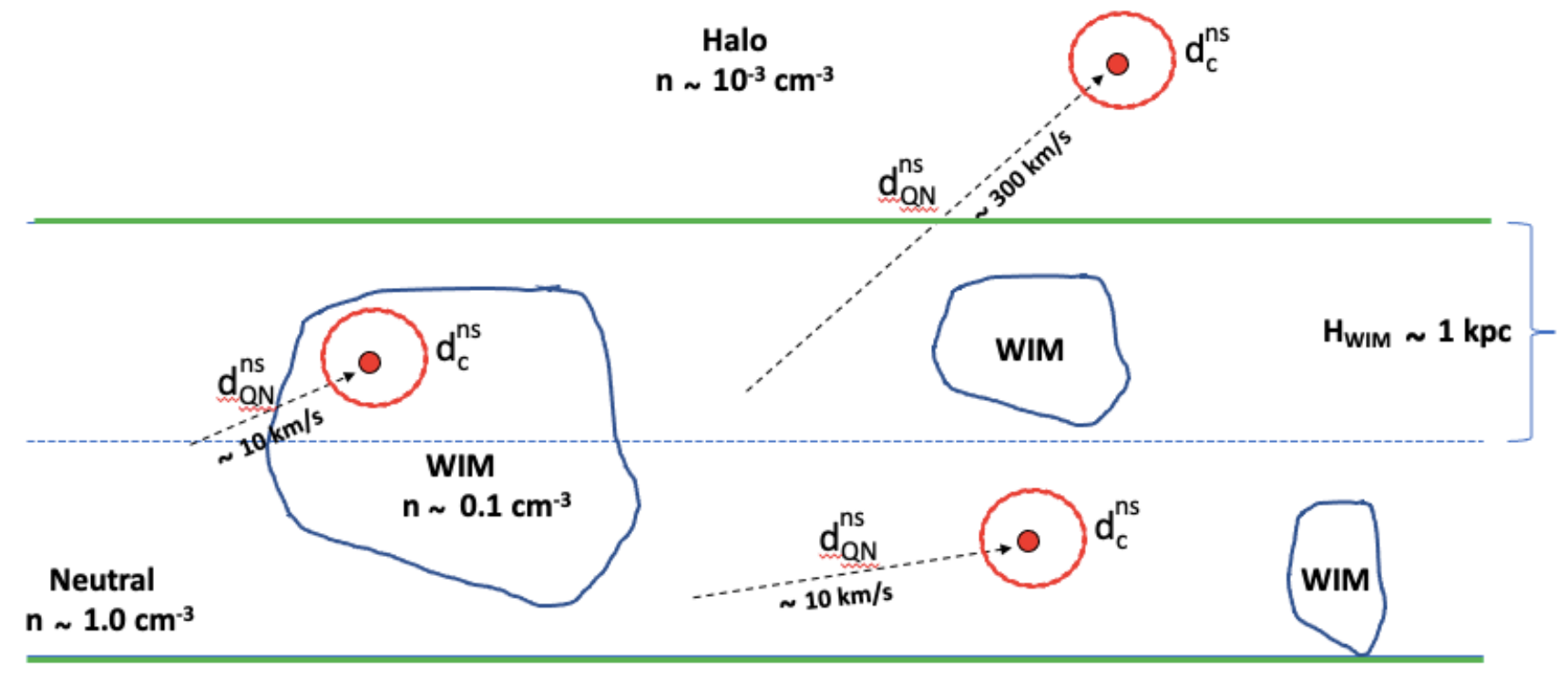}
\caption{{\bf The birthplace and burst environment of QN chunks}:
This illustration showcases various scenarios involving different media both inside and outside a galaxy where a QN event can occur, leading to the transition of chunks into a collisionless state. Eventually, these chunks undergo the BI-WI instabilities, resulting in CSE bursts. During a QN event, millions of chunks are ejected, expanding radially away from the QN source. The resulting FRBs can be classified into three primary scenarios: (i) Halo-Born-Halo-Bursting chunks, which produce one-off FRBs ($R_{\rm halo}<\Delta r_{\rm BI-WI}^{\rm ns}$); 
(ii) Disk-Born-Halo-Bursting chunks, which yield radially repeating FRBs ($R_{\rm halo}>>\Delta r_{\rm BI-WI}^{\rm ns}$);
(iii) Disk-Born-Disk-Bursting chunks, leading to erratically repeating FRBs.
 Angular repetition can occur in any of these three scenarios when peripheral chunks from the same QN event are detected.}
 \label{fig:WIM-and-Halo}
\end{figure*}

\subsection{CSE luminosity}
\label{appendix:CSE-luminosity}

For each ambient electron swept-up by the chunk, a fraction $\zeta_{\rm BI}$ of its kinetic energy is first converted by the BI into heat which 
is  subsequently emitted as CSE by the collisionless chunk's accelerated electrons (via the WI). The CSE luminosity  is 
 $L_{\rm CSE}= (\zeta_{\rm BI}\Gamma_{\rm c}m_{\rm e}c^2)\times (\pi R_{\rm c}^2 \Gamma_{\rm c}n_{\rm amb.}^{\rm ns}c)$. Or,
 \begin{equation}
 \label{eq:LCSE}
L_{\rm CSE} \simeq  
6.2\times 10^{36}\ {\rm erg\ s}^{-1}\times \zeta_{\rm BI}\Gamma_{\rm 2.5}^{2/5} m_{\rm c, 22.3}^{3/5}n_{\rm amb., -1}^{\rm ns\ 1/5}\ ,
\end{equation}
where we made use of Eq. (\ref{eq:ncandRc}) and recall that here $n_{\rm amb.}^{\rm ns}$ is in the NS frame. 
 The kinetic energy of the chunk converted to CSE emission is $E_{\rm CSE}=L_{\rm CSE} t_{\rm BI, s}$ with 
$t_{\rm BI, s}\simeq \frac{3}{2}(m_{\rm p}/m_{\rm e})^{1/3}/\nu_{\rm p, e}$  the BI saturation time (see Appendix SC in \citealt{ouyed2021}). 
Thus

\begin{equation}
 \label{eq:ECSE}
E_{\rm CSE} \simeq  
1.1\times 10^{38}\ {\rm erg\ s}^{-1}\times \frac{\zeta_{\rm BI}\Gamma_{\rm 2.5}^{2/5} m_{\rm c, 22.3}^{3/5}n_{\rm ionized, -1}^{\rm ns\ 1/5}}{\nu_{\rm p, e}}\ .
\end{equation}

We should point out that $n_{\rm amb.}^{\rm ns}$ here is the density of the ionized plasma the chunk is propagating in;
otherwise no BI and WI can occur.
This density may be different from that of the medium (neutral or ionized) where the chunk first became collisionless
and which sets and fixes the chunk's intrinsic density $n_{\rm c}$ and thus its plasma frequency and related quantities.
This is a fundamental difference as illustrated in Figure \ref{fig:WIM-and-Halo} but which has an impact only on the luminosity
(i.e. the FRB fluence) during bursting.  

When considering the two environments with $n_{\rm ionized}^{\rm ns}\neq n_{\rm amb.}^{\rm ns}$, 
there are  three main scenarios 
for the resulting FRBs as illustrated in Figure \ref{fig:WIM-and-Halo}:
(a) Halo-Born-Halo-Bursting chunks (yielding one-off FRBs);
(b) Disk-Born-Halo-Bursting chunks (yielding  radially repeating FRBs);
(c) Disk-Born-Disk-Bursting chunks (yielding radial FRBs with erratic repetition).

Hereafter we assume  that the bursting environment is the birth environment (i.e. $n_{\rm ionized}^{\rm ns}=n_{\rm amb.}^{\rm ns}$) but one should keep in mind that if and when necessary one must multiply the CSE luminosity
by $n_{\rm ionized}^{\rm ns}/n_{\rm amb.}^{\rm ns}$.  With $n_{\rm ionized}^{\rm ns}=n_{\rm amb.}^{\rm ns}$ we can express $E_{\rm CSE}$ in terms of $\nu_{\rm p, e}$ via $n_{\rm amb.}^{\rm ns}\propto n_{\rm c}^{5/6}\propto \nu_{\rm p, e}^{5/3}$.

To summarize, the key equations (in the chunk's frame) defining the CSE in our model are 

\begin{align}
\label{eq:chunk-frame}
\nu_{\rm CSE, p, min.}&=\nu_{\rm p, e}\\\nonumber
\nu_{\rm CSE, p, max.}&= \frac{\nu_{\rm p, e}}{\beta_{\rm WI}^{1/2}}\\\nonumber
t_{\rm m-WI}&=  \sqrt{\frac{m_{\rm p}}{m_{\rm e}}} \times \frac{\zeta_{\rm m-WI}}{\nu_{\rm p, e}}\\\nonumber
E_{\rm CSE}&\simeq 1.4\times 10^{36}\ {\rm ergs}\times \frac{\zeta_{\rm BI}m_{\rm c, 22.3}^{7/12}}{\nu_{\rm p, e}^{2/3}}\ ,
\end{align}
which all depend on the chunk's plasma frequency $\nu_{\rm p, e}$. The CSE duration is given in Eq. (\ref{eq:dt-CSE}).
Because $E_{\rm CSE}<<(E_{\rm c}^{\rm ke})^{\rm ns}=\frac{m_{\rm e}}{m_{\rm p}}\times \Gamma_{\rm c} m_{\rm c}c^2$, a chunk can experience many CSE episodes during its lifetime (see Appendix \ref{appendix:radial-periodicity}).

\renewcommand\thesection{\Alph{section}}
\renewcommand{\thefigure}{\thesection.\arabic{figure}}
\setcounter{figure}{0} 

\section{Radial repetition}
\label{appendix:radial-periodicity}

Once the Weibel instability saturates, electrons will be trapped by the generated magnetic field. Coulomb collisions set-in and the
dissipation of the magnetic field starts. Heating of electrons during filament merging ceases once the maximum filament size is reached at which point CSE cools electrons to $\sim$ 511 keV.  The diffusive decay timescale of the WI-generated magnetic field can be estimated from
$t_{\rm B, diff.}\sim \lambda_{\rm F, max.}^2/\eta_{\rm B}$ where $\lambda_{\rm F, max.}\sim c/\nu_{\rm p, e}$ is the filament's maximum size and
$\eta_{\rm B}= m_{\rm e} c^2 \nu_{\rm coll.}/4\pi n_{\rm c, e}e^2 = \nu_{\rm coll.}c^2/\nu_{\rm p, e}^2$ the diffusion coefficient with $n_{\rm c, e}=n_{\rm c}$ (e.g. \citealt{lang1999}).
 The electron Coulomb collision frequency  is $\nu_{\rm coll.}=9.2\times 10^{-11}\ {\rm s}^{-1}\times n_{\rm c, 0}\ln{\Lambda}/T_{\rm e, keV}^{3/2}$ where the electron temperature is in keV (e.g. \citealt{richardson2019}). 
A more rigorous assessment of the dissipation of the magnetic field requires simulations and is beyond the scope of this paper.
However, it is reasonable to take the analytical value $t_{\rm B, diff.}\sim  1/\nu_{\rm coll.}$ as an order of magnitude estimate. 
 
A maximum dissipation timescale can be estimated by taking $T_{\rm e}=511$ keV and $\ln{\Lambda}=10$ to get 
  $t_{\rm B, diff.}\sim 1.7\times 10^6\ {\rm years}/n_{\rm c, 0}$. It is shorter if for example the electrons are cooled below 511 keV.
  We take this into account by introducing a new parameter $\alpha_{\rm diff.}<1$ such that 
   \begin{equation}
 \label{eq:tBdiff}
 t_{\rm B, diff.}\sim \alpha_{\rm diff.}\times  \frac{1.7\times 10^6\ {\rm years}}{n_{\rm c, 0}}\ .
 \end{equation} 

Once the magnetic field dissipates, another BI-WI is triggered and a new CSE episode starts as the chunk keeps 
 propagating in the ambient plasma. 
In terms of the plasma frequency (with $n_{\rm c}\propto \nu_{\rm p, e}^2$), and in the observer's frame, the corresponding time period between BI-WI episodes is $\Delta t_{\rm BI-WI}^{\rm obs.}= (1+z)t_{\rm B, diff.}/D(\Gamma_{\rm c},\theta_{\rm c})$. I.e.

\begin{align}
\label{eq:Pcrad}
\Delta t_{\rm BI-WI}^{\rm obs.}&\sim 31.8\ {\rm days}\times \frac{\alpha_{\rm diff.}}{(1+z)f(\Gamma_{\rm c},\theta_{\rm c})}\times \frac{\Gamma_{\rm c, 2.5}}{{\nu_{\rm p, e, 9}^{\rm obs.}}^2}\\\nonumber
&\sim 267.8\ {\rm days}\times  (1+z)f(\Gamma_{\rm c},\theta_{\rm c})\times \frac{\alpha_{\rm diff.}}{\Gamma_{\rm c, 2.5}^{17/5}m_{\rm c, 22.3}^{1/10} n_{\rm amb., -1}^{\rm ns\ 6/5}}\ .
\end{align}

The  distance travelled between successive BI-WI episodes, $\Delta r_{\rm BI-WI}^{\rm ns}= c\Gamma_{\rm c} t_{\rm B, diff.}$ (in the NS frame), is

\begin{align}
\label{eq:distance}
\Delta r_{\rm BI-WI}^{\rm ns}&\sim 28.7\ {\rm kpc}\times \frac{\alpha_{\rm diff.}}{\Gamma_{\rm c, 2.5}^{7/5}m_{\rm c, 22.3}^{1/10}n_{\rm amb., -1}^{\rm ns\ 6/5}}\ .
\end{align}

Radial repetition would ensue if the length of the ionized ambient plasma, $L_{\rm ionized}$, 
exceeds $\Delta r_{\rm BI-WI}^{\rm ns}$. This translates to
\begin{equation}
\label{eq:nambc}
n_{\rm amb.}^{\rm ns} > n_{\rm amb., c}^{\rm ns}\sim 0.31\ {\rm cm}^{-3}\times \frac{\alpha_{\rm diff.}^{5/6}}{L_{\rm ionized, 10 {\rm kpc}}^{5/6}\Gamma_{\rm c, 2.5}^{7/6}m_{\rm c, 22.3}^{1/12}} \ ,
\end{equation}
which is very weakly dependent on the chunk's mass $m_{\rm c}$. 
The  higher the $n_{\rm c}$ the higher the plasma frequency (i.e. a higher Coulomb collision frequency) and
thus the faster the magnetic dissipation timescale $\propto n_{\rm c}^{-1}$ which translates to shorter 
time intervals $\Delta t_{\rm BI-WI}^{\rm obs.}$ between BI-WI episodes.

 \renewcommand\thesection{\Alph{section}}
\renewcommand{\thefigure}{\thesection.\arabic{figure}}
\setcounter{figure}{0}

\section{The spectrum}
\label{appendix:spectrum}

In our model, the CSE spectrum from a given chunk is a convolution of the spectrum from a single bunch of size
$\lambda_{\rm b}$ (with $\nu_{\rm p, e} < \nu_{\rm CSE} \le \nu_{\rm CSE, p}(t)=\frac{c}{\lambda_{\rm b}(t)}$) and
the distribution of bunch size $\lambda_{\rm b}$ within a given chunk.  The distribution of bunch size
and its evolution as the filaments merge remains to be determined. 
The temporal evolution of the convolved spectrum depends on the filament merging rate. I.e. as the
filaments grow in size so does the peak of the distribution of bunch size which translates to the decrease in 
the overall maximum CSE frequency.

\begin{figure*} 
\centering
\includegraphics[scale=0.4]{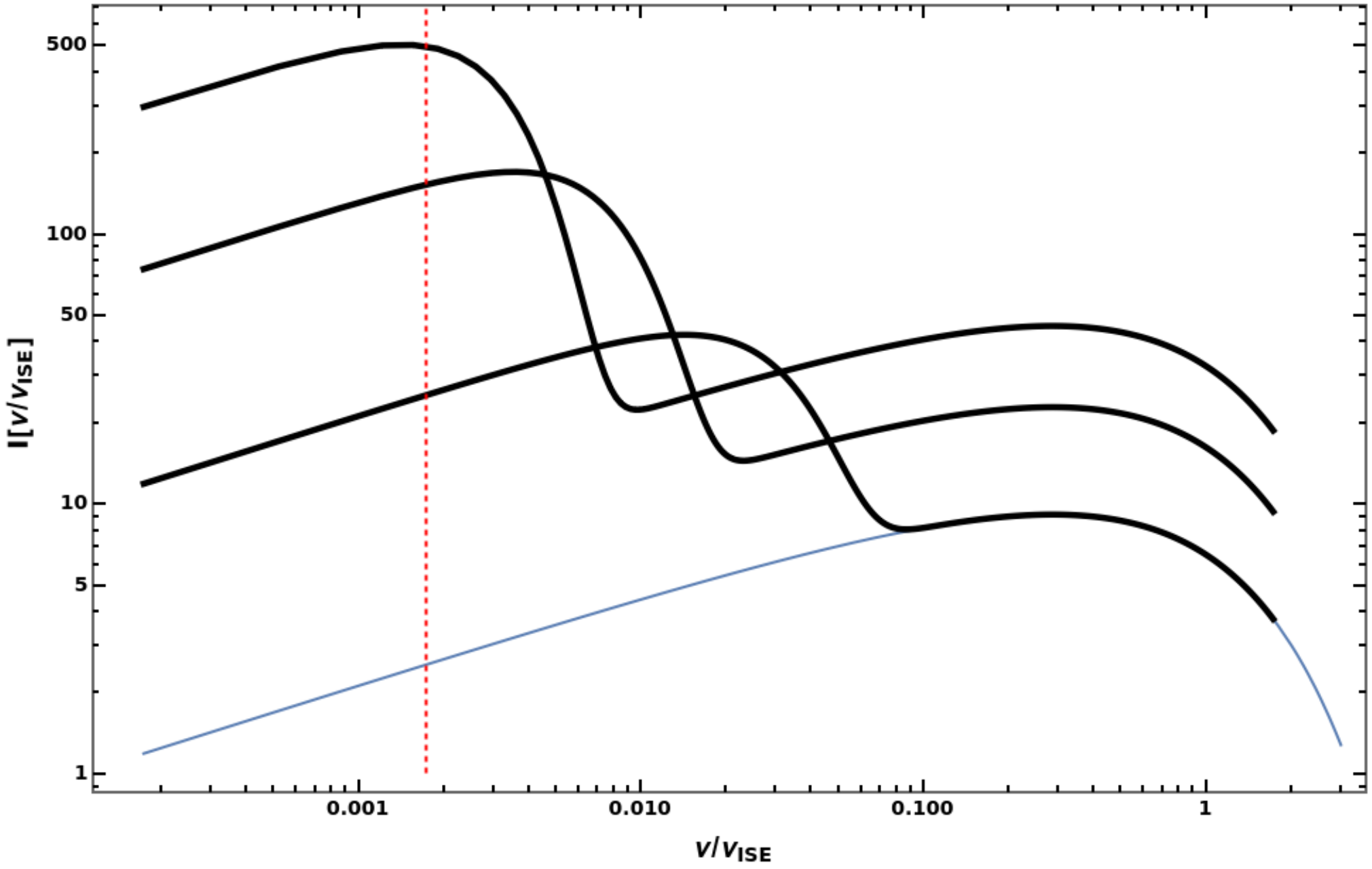}
\caption{{\bf Normalized CSE spectrum}: Normalized intensity  versus $\nu / \nu_{\rm ISE, p}$  with 
 the incoherent synchrotron peak frequency $\nu_{\rm ISE, p}=\frac{3}{2}\gamma_{\rm e}^2\nu_{\rm B}$;  $\nu_{\rm B}=\sqrt{ \frac{m_{\rm p}}{m_{\rm e}}} \nu_{\rm p, e}$ in our model. The red vertical line shows the chunk's plasma frequency $\zeta_{\rm p, e}=\nu_{\rm p,e}/\nu_{\rm ISE, p}=(\frac{3}{2}\gamma_{\rm e}^2\sqrt{\frac{m_{\rm p}}{m_{\rm e}}})^{-1}$.  The bottom curve is the
incoherent synchrotron spectrum in the  $\nu_{\rm p, e} < \nu < \nu_{\rm CSE, p}$ frequency range ($\zeta_{\rm p, e} <\nu / \nu_{\rm ISE, p}< 1$) from a single bunch containing $N_{\rm e}=20$ electrons
and $\gamma_{\rm e}=3.0$  (see Eq. (\ref{appendix:single-electron-spectrum})).  The corresponding CSE spectrum
with $(\nu_{\rm CSE, p},N_{\rm e})= (20\nu_{\rm p, e}, 10)$ is also shown which follows the $\nu^{1/3}$ of ISE curve at $\nu<\nu_{\rm CSE, p}$ and
a steep exponential decay $e^{-(\nu_{\rm CSE, p})^2}$ for $\nu>\nu_{\rm CSE, p}$.  The two top curves depict the time evolution
of CSE intensity from the bunch as it grow in size with $(\nu_{\rm CSE, p},N_{\rm e})= (5\nu_{\rm p, e}, 25)$ and $(2\nu_{\rm p, e}, 50)$, respectively.
As the bunch grows in size, $\nu_{\rm CSE, p}=c/\lambda_{\rm b}$ decreases while the number of electrons $N_{\rm e}$ increases;
see Appendix \ref{appendix:spectrum} for details.
 }
\label{fig:CSE-spectrum}
\end{figure*}

\subsection{A single bunch}
 
 We start with the incoherent synchrotron emission spectrum (ISE) for a single electron with Lorentz factor $\gamma_{\rm e}$ (e.g. \citet{jackson1999})

\begin{equation}
\label{appendix:single-electron-spectrum}
F(\zeta) = \frac{9}{\sqrt{3}{8\pi} }\zeta \int_{\zeta}^\infty K_{5/3}(x) dx\ ,
\end{equation}
with $\zeta=\nu/\nu_{\rm ISE, p}$ and $\nu_{\rm ISE, p}= \frac{3}{2}\gamma_{\rm e}^2\nu_{\rm c}$ the incoherent spectrum peak frequency while $\nu_{\rm c}$ is the cyclotron frequency.  The $K_{5/3}$ is the modified Bessel function of second kind of order  5/3. Due to the magnetic field reaching near equipartition with the chunk's protons at WI saturation we have 
 $\nu_{\rm c}\sim \sqrt{ \frac{m_{\rm p}}{m_{\rm e}}}\nu_{\rm p, e}$ .
 In these units, 
$\zeta_{\rm p, e}=\nu_{\rm p, e}/\nu_{\rm ISE}= (\frac{3}{2}\gamma_{\rm e}^2\sqrt{\frac{m_{\rm p}}{m_{\rm e}}})^{-1}$.

For simplicity, we assume a mono-energetic electron population per bunch. 
For a bunch which contains $N_{\rm e}$ electrons and which has a Gaussian distribution and r.m.s bunch length $\lambda_{\rm b}$ (with $\nu_{\rm CSE, p}=\frac{c}{\lambda_{\rm b}}$), the spectrum  is (see Appendix SD in \citet{ouyed2020} and references therein) 

\begin{equation}
F_{\rm b}(\zeta, \zeta_{\rm CSE, p}(t)) = F(\zeta)\times N_{\rm e}\left( 1+(N_{\rm e}-1) e^{-\left(\frac{\zeta}{\zeta_{\rm CSE, p}(t)}\right)^2}\right)\ ,
\end{equation}
with $\zeta_{\rm CSE, p}(t)=\nu_{\rm CSE, p}(t)/\nu_{\rm ISE}$.  Also, $\zeta_{\rm CSE, p, max.}= \zeta_{\rm p, e}/\beta_{\rm WI}^{1/2}$ and $\zeta_{\rm CSE, p, min.}= \zeta_{\rm p, e}$.  

The $F(\zeta)\times N_{\rm e}$ term is the ISE intensity
which scales linearly with $N_{\rm e}$ in the  $\zeta_{\rm p, e}\le \zeta \le \zeta_{\rm ISE, p}$ frequency range.
The CSE dominates  when (using $N_{\rm e}-1\simeq N_{\rm e}$)
\begin{equation}
N_{\rm e} e^{-\left(\frac{\zeta}{\zeta_{\rm CSE, p}}\right)^2} > 1\quad {\rm or}\quad \nu < \nu_{\rm CSE, p}\times \frac{\sqrt{\ln{N_{\rm e}}}}{2\pi}\sim
\nu_{\rm CSE, p} \ ,
\end{equation}
with $N_{\rm e}$ given in Eq. (\ref{eq:Ne}). I.e. for $\nu < \nu_{\rm CSE, p}$ the spectrum is the $N_{\rm e} F(\zeta)$
spectrum boosted in intensity by $N_{\rm e}$. The CSE follows the $\nu^{1/3}$ slope for $\nu<\nu_{\rm CSE, p}$
 while it decays exponentially as $e^{-(\frac{\nu}{\nu_{\rm CSE, p}})^2}$ for $\nu>\nu_{\rm CSE, p}$.

 In Figure \ref{fig:CSE-spectrum}, we show an example of CSE spectrum for $\gamma_{\rm e}=3$
 and three different time steps with $(\nu_{\rm CSE, p},N_{\rm e})= (20\nu_{\rm p, e}, 10)$,
  $(5\nu_{\rm p, e}, 25)$ and $(2\nu_{\rm p, e}, 50)$.
  Over time, the peak CSE frequency decreases (due to bunch size growth) as $\nu_{\rm CSE, p}\propto t^{-\delta_{\rm m-WI}}$ while the number of electrons in a bunch increases as $N_{\rm e}\propto t^{2\delta_{\rm m-WI}}$ boosting the CSE.

\subsection{A distribution of bunches}

Consider $N_{\rm b}$ bunches with a distribution in size $\lambda_{\rm b}$ and a corresponding $\nu_{\rm CSE, p}=c/\lambda_{\rm b}$
distribution ${\rm PDE}(\nu_{\rm CSE, p})$
where ${\nu_{\rm CSE, p, min.}}=c/\lambda_{\rm b, min.} \le \nu_{\rm CSE, p} \le {\nu_{\rm CSE, p, max.}}^{\rm max.}=c/\lambda_{\rm b, max.}$. 
The cumulative CSE spectrum is then

\begin{equation}
F_{\rm cum.}(\zeta) = \int_{ {\zeta_{\rm CSE, p, min.}}}^{{\zeta_{\rm CSE, p, max.}}}F_{\rm b}(\zeta,\zeta_{\rm CSE, p}) {\rm PDE}(\zeta_{\rm CSE, p}) d\zeta_{\rm CSE, p}\ .
\end{equation}
So far we ignored the turbulent nature of the filament merging during the WI and the intermittent acceleration of the
chunk electrons. The final spectrum is much more complex than what is presented here and 
will be studied elsewhere.

 \end{appendix}

\end{document}